\RequirePackage{amsmath}
\documentclass[traditabstract]{aa} 
\usepackage{natbib}
\usepackage{rotating}
\usepackage{graphicx}
\usepackage{amsmath}
\usepackage[switch]{lineno}
\usepackage{txfonts}
\usepackage{amssymb}
\begin{document} 

   \title{The flat-spectrum radio quasar 3C 345 from the high to the low emission state}


   \author{M. Berton
	\inst{1,2}\thanks{marco.berton@unipd.it}
	\and N.~H. Liao\inst{3}
	\and G. La Mura\inst{1}
	\and E. J{\"a}rvel{\"a}\inst{4,5}
	\and E. Congiu\inst{1,2}
	\and L. Foschini\inst{2}
	\and M. Frezzato\inst{1}
	\and	\\
	V. Ramakrishnan\inst{6}
	\and X.~L. Fan\inst{7,8}
	\and A. L\"ahteenm\"aki\inst{4,5}	
	\and T. Pursimo\inst{8}
	\and V. Abate\inst{1}
	\and J.~M. Bai\inst{7}
	\and\\
	P. Calcidese\inst{9}
	\and S. Ciroi\inst{1}
	\and L. Chen\inst{10}
 	\and V. Cracco\inst{1}
	\and S.~K. Li\inst{7}
	\and M. Tornikoski\inst{4}
	\and P. Rafanelli\inst{1}
         }

   \institute{$^{1}$ Dipartimento di Fisica e Astronomia "G. Galilei", Universit\`a di Padova, Vicolo dell'Osservatorio 3, 35122 Padova, Italy;\\
   $^{2}$ INAF - Osservatorio Astronomico di Brera, via E. Bianchi 46, 23807 Merate (LC), Italy;\\
   $^{3}$ Purple Mountain Observatory, Chinese Academy of Sciences, 2 West Beijing Road, Nanjing 210008, China; \\
   $^{4}$ Aalto University Mets\"ahovi Radio Observatory, Mets\"ahovintie 114, 02540 Kylm\"al\"a, Finland; \\
   $^{5}$ Aalto University Department of Radio Science and Engineering, P.O. BOX 13000, FI-00076 AALTO, Finland; \\
   $^{6}$ Universidad de Concepci\'on, Departamento de Astronom\'ia, Casilla 160-C, Concepci\'on, Chile; \\
 $^{7}$ Yunnan Observatories, Chinese Academy of Sciences, Kunming, Yunnan 650011, China; \\
 $^{8}$ School of Physics, Huazhong University of Science and Technology, Wuhan 430074, China; \\
 $^{9}$ Nordic Optical Telescope, Apartado 474, 38700 Santa Cruz de La Palma, Spain;\\
 $^{10}$ Osservatorio Astronomico della Regione Autonoma Valle d'Aosta, Saint Barthelemy, I-11020 Nus, Italy;\\
 $^{11}$ Shanghai Astronomical Observatory, Chinese Academy of Sciences, 80 Nandan Road, Shanghai 200030, China. 
             }

\authorrunning{M. Berton et al.}
\titlerunning{High and low states of 3C 345}

\abstract{We report simultaneous observations at different energy bands in radio, optical, UV, X-rays and $\gamma$ rays of the flat-spectrum radio-quasar 3C 345. We built the light curve of the source at different frequencies from 2008, the beginning of the \textit{Fermi} all-sky survey, to 2016, using new data and public archives. In particular we obtained several optical spectra, to study the behavior of emission lines and the continuum in different activity states and to derive the black hole mass. 3C 345 showed two flaring episodes in 2009, which occurred simultaneously in $\gamma$ ray, optical/UV and X-rays, and were later followed in radio. The source shows an inverse Compton dominated spectral energy distribution, which moved from higher to lower frequencies from the high to the low state. The reverberation of emission lines during one outburst event allowed us to constrain the location of production of $\gamma$ rays very close to the broad-line region, and possibly in the jet-base. We report the observation of an increased accretion after the outburst, possibly induced by the decrease of magnetic field intensity with respect to the low state.}

\maketitle

\newcommand{\kms}{km s$^{-1}$}
\newcommand{\ergs}{erg s$^{-1}$}
\section{Introduction}
Active galactic nuclei (AGN) are among the most luminous sources in the Universe. The source of their powerful emission throughout the whole electromagnetic spectrum is accretion onto a supermassive black hole, with a mass usually ranging between 10$^5$ and 10$^{10}$ M$_\odot$. This central engine is often able to launch a relativistic jet, which can carry energy and momentum far away from the nucleus, emitting a large amount of radiation at all frequencies. When the jet is pointed toward the Earth, the relativistic beaming enhances the source luminosity, making it a blazar. Such sources are often associated with $\gamma$-ray emission, and they are here divided into three classes according to the properties of their optical spectrum and their overall spectral energy distribution (SED): BL Lacertae objects (BL Lacs), flat-spectrum radio quasars (FSRQ) and flat-spectrum radio-loud narrow-line Seyfert 1 galaxies (F-NLS1, \citealp[e.g.][]{Foschini17}). While the optical spectrum of BL Lacs is basically a featureless continuum, the other two classes of blazars are characterized by the presence of strong emission lines, and are thought to be AGN powered by a radiatively efficient accretion mode \citep{Heckman14}. Flat-spectrum radio quasars show usually permitted lines with a high full width at half maximum (FWHM), and a large black hole mass, above 10$^8$ M$_\odot$. Conversely, F-NLS1 have a FWHM(H$\beta$) by definition lower than 2000 \kms, and might be a young evolutionary phase of FSRQs \citep{Foschini15, Berton16c}. \par
The spectral energy distribution (SED) of blazars is approximatively characterized by two humps. The low frequency part of the SED originates via synchrotron radiation, while the second peak is due to inverse Compton scattering (IC) of relativistic electrons on seed photons. The beamed jet also makes blazars extremely variable throughout all frequencies. In particular they can rapidly move from a high state, in which their luminosity is strongly increased, to a low state, in which the jet contribution is less pronounced and the luminosity usually decreases. To better understand the changes that occur between these states of blazar activity, and how these changes reflect in the optical spectral lines, we decided to focus on 3C 345, a FSRQ at z = 0.593 \citep{Lynds65}, with coordinates R.A. 16h42m58.8s and Dec. +39d48m37.2s (J2000). \par
This quasar is part of a Radio AND Optical Monitoring (RANDOM) campaign, which investigates a small sample of six line-emitting AGN, selected because of their strong optical and radio emission. The monitoring has been carried out since February 2016 with the Mets{\"a}hovi Observatory in the radio band at 37 GHz and in the optical band with the Asiago 1.22m telescope. In the future, polarimetric and spectropolarimetric observations will be also carried out with the Copernico 1.82m telescope of the Istituto Nazionale di Astrofisica (INAF)/Asiago Observatory. \par
The radio quasar 3C 345 is a well studied object showing superluminal motion on very large baseline interferometry (VLBI) scales (\citealp[see the MOJAVE project,][see also]{Lister09} \citealp{Roberts12, Roberts13}). Its optical spectrum, as expected from a FSRQ, is that of a type 1 AGN although X-ray observations suggest the presence of a Compton thick absorber in its nucleus similar to those found in type 2 AGN \citep{Eguchi17}. The first optical magnitude measurements were already carried out in the late nineteenth century, and in the following years the source showed an intense flaring activity \citep{Zhang98}. Some authors suggested that such flaring episodes are induced by perturbations on the accretion flow and disk generated by the presence of a binary system of supermassive black holes \citep{Lobanov05, Klare05}. A possible consequence of these is the complex structure of the relativistic jet, which exhibits an helical parsec-scale shape \citep{Schinzel10}, and a counter-jet symmetric to the approaching jet \citep{Matveyenko13}. The jet remains relativistic for hundreds of parsecs, and it is also a strong X-ray emitter \citep{Kharb12}. Radio quasar 3C 345 has also been identified as a $\gamma$-ray emitter by the \textit{Fermi LAT} collaboration \citep[e.g.][]{Schinzel11}, and in the third \textit{FERMI LAT} source catalog (3FGL, \citealp{Ackermann15}) it was associated with the source 3FGL J1642.9+3950. During the monitored period, the source exhibited three strong $\gamma$-ray flares that we refer to as 2009A, 2009B, and 2015 (see Sect.~3), before entering a low activity state that extended for the rest of the observational campaign described here.\par
\begin{figure}[t!]
\centering
\includegraphics[width=\hsize]{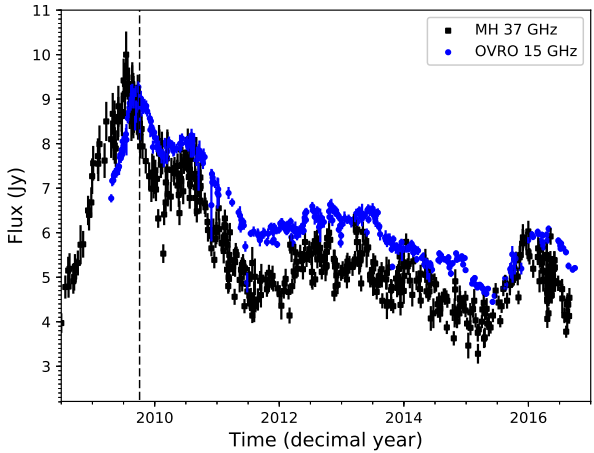} 
\caption{Light curve of 3C 345 in radio. The black points indicate the flux at 37 GHz, measured by the Mets\"ahovi Radio Telescope. The blue points indicate the flux at 15 GHz, measured by the Owen Valley Radio Telescope. The vertical dashed line indicates the 2009B flare. }
\label{fig:curva_radio}
\end{figure}

\begin{figure*}[t!]
\centering
\includegraphics[trim={3cm 0cm 3cm 0},clip,width=\textwidth]{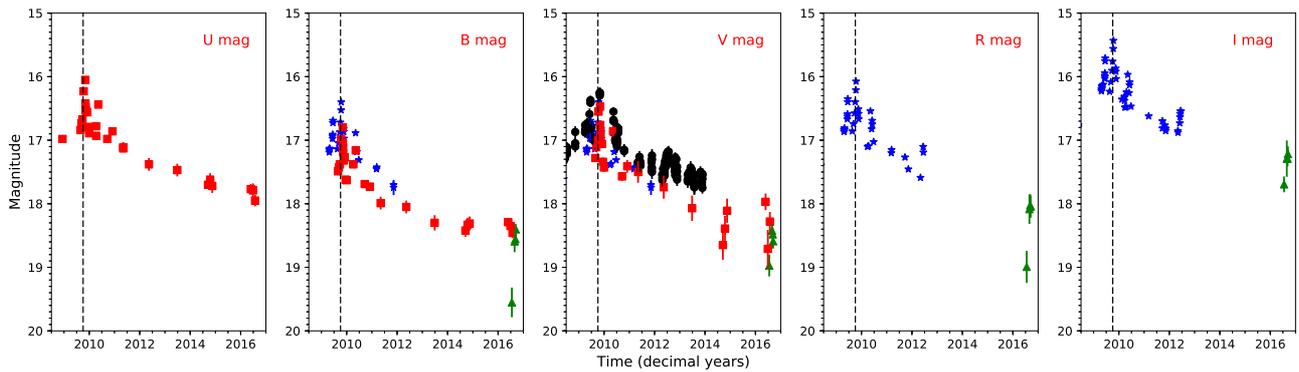} 
\caption{Light curve of 3C 345 in the optical filters. The black circles are those measured by CSS. Red squares come from Swift/UVOT, blue stars from the Kunming and Lijiang telescopes, the green triangles are measured by the Asiago Schmidt telescope. From left to right, U magnitude, B magnitude, V magnitude, R magnitude, and I magnitude. The vertical dashed line indicates the 2009B flare.}
\label{fig:curva_ottico}
\end{figure*}

\begin{table*}[t!]
\caption{Observational details for optical spectra.}
\label{tab:spettri}
\centering
\begin{tabular}{l c c c c c c}
\hline\hline
Date & Telescope & Dispersion & Slit (arcsec) & R & Lamp & Exposure time (s) \\
\hline
2009-08-20 & NOT & grism\#6 & 1.3 & 530 & HeNe & 700 \\
2010-06-30 & NOT & grism\#4 & 1.0 & 400 & HeNe & 900 \\
2011-05-29 & NOT & grism\#4 & 1.0 & 400 & HeNe & 500 \\
2013-07-16 & INT & grating\#7 & 1.0 & 400 & CuArNe & 1800 \\
2016-05-26 & NOT & grism\#20 & 0.5 & 1850 & ThAr & 1800 \\
2016-05-30 & TNG & LR-B & 0.7 & 750 & HgNe & 1800 \\
2016-08-28 & T122 & 300 mm$^{-1}$ & 4 & 700 & FeAr & 14400 \\
\hline
\end{tabular}
\tablefoot{Columns: (1) Date; (2) telescope used for observations; (3) dispersion element; (4) slit; (5) spectral resolution R; (6) lamp for wavelength calibration; (7) exposure time.}
\end{table*}

In radio the source is bright and reveals a complex morphology as shown by the MOJAVE blazar monitoring observations at 15 GHz \citep{Lister09}. The parsec-scale emission has the same long-term behavior as optical and $\gamma$-ray radiation \citep{Schinzel12}. On larger scale, the source has been monitored since the 1980s from Mets{\"a}hovi at 37 GHz. Both in radio and in optical bands, 3C 345 flux has been showing a decreasing trend, in particular since the last major flare in 2009 (see Fig.~\ref{fig:curva_radio}), with some minor flaring events. In 2016 the source reached in the optical band a minimum since the beginning of the observations. Therefore we decided to study via long-term multiwavelength observations the evolution of the source from the last major flare to this very low state. We retrieved archival data and performed new observations between 2008 August 4, the beginning of \textit{Fermi} observations, and 2016 August 31. We built the spectral energy distribution (SED) of the source to derive its main physical properties, and we studied optical spectrum and emission lines to better constrain its characteristics. \par
In Sect.~2 we describe the multiwavelength data analysis, in Sect.~3 we show our results, in Sect.~4 we discuss them, and in Sect.~5 we present a brief summary. Throughout this work we adopt a standard $\Lambda$CDM cosmology, with a Hubble constant $H_0 = 70$ \kms\ Mpc$^{-1}$, $\Omega_m = 0.27$, and $\Omega_\Lambda = 0.73$ \citep{Komatsu11}. 
\section{Observations and data analysis}

\subsection{Radio band} 
The 13.7-metre radio telescope at the Aalto University Mets\"ahovi Radio Observatory\footnote{http://www.metsahovi.fi/} in Finland is used for monitoring large samples of AGN at 37 GHz. The measurements are made with a 1 GHz-band dual beam receiver centered at 36.8 GHz. The telescope is equipped with a high electron mobility pseudomorphic transistor (HEMPT) front end that operates at room temperature. The observations are on-on observations, alternating the source and the sky in each feed horn. A typical integration time to obtain one flux density data point is between 1200 and 1400 s. The detection limit of the telescope at 37 GHz is on the order of 0.2 Jy under optimal conditions. Data points with a signal-to-noise ratio $<$ 4 are handled as non-detections. The flux density scale is set by observations of DR 21. Sources NGC 7027, M 87 (3C 274), and NGC 1275 (3C 84) are used as secondary calibrators. A detailed description of the data reduction and analysis is given by \citet{Teraesranta98}. The error estimate in the flux density includes the contribution from the measurement root mean square and the uncertainty of the absolute calibration. The monitoring of 3C 345 at 37 GHz has been carried out systematically with Mets\"ahovi since 1980. The light curve is indicated by the black squares in Fig.~\ref{fig:curva_radio}. \par
We also retrieved the public data obtained with the 40m telescope of the Owen Valley Radio Observatory (OVRO). 
The 389 observations were carried out at 15 GHz, and they were part of the \textit{Fermi} blazar monitoring program \citep{Richards11}. 
The light curve is represented by the blue circles in Fig.~\ref{fig:curva_radio}. 
\subsection{Optical}
\subsubsection{Photometry}
During the low state, we obtained a magnitude estimate for g, r, and i Sloan Digital Sky Survey (SDSS) filters using the 1.82m Copernico Telescope of the INAF/Asiago Observatory.\footnote{http://archive.oapd.inaf.it/asiago/} This observation was performed on 2016 April 16, with a seeing of 2", in 2x2 binning and with a partially overcast sky. For each band we got an exposure of 180 s. The source was observed again on 2016 June 16 and 2016 June 25 with the Schmidt 92/67cm of the INAF/Asiago Observatory. The target was observed using B, V, R, and I filters, getting three exposures of 300s each for all filters. The source was also observed by the Nordic Optical Telescope (NOT)\footnote{http://www.not.iac.es/} in four different epochs, and by the Isaac Newton Telescope (INT)\footnote{http://www.ing.iac.es/Astronomy/telescopes/int/} in 2013. All these images were processed using the standard reduction, correcting for bias, dark (only for the Schmidt 92/67cm data), and flat-field. To obtain the magnitudes we used a point spread function (PSF) fitting technique, which can be used for point-like sources. The images were astrometrized finding the surrounding field stars in the Two Micron All-Sky Survey (2MASS) archive. The instrumental magnitudes were then converted into physical quantities using the corresponding SDSS magnitudes. \par
To better constrain the historical behavior of 3C 345 in optical, we also used the photometric observations performed with the Lijiang 2.4m telescope and the Kunming 1.02 m telescope at the Yunnan Observatories.\footnote{http://www.gmg.org.cn/english/} Before 2008 July 30 a PI 1024TKB CCD 1024x1024 px was mounted at the Cassegrain focus of Kunming 1.02 m telescope, later replaced by an Andor DW436 CCD with 2048x2048 px. At the Lijiang 2.4m telescope a PIVersArray 1300B CCD was equipped at the Cassegrain focus until 2011. Later the observations were carried out by the Yunnan Faint Object Spectrograph and Camera (YFOSC). The observations were performed with different exposure times, to match the different seeing and weather conditions. The typical seeing was $<$2.5" at Kunming telescope and $<$2" at the Lijiang telescope. The data were reduced following the standard Image Reduction and Analysis Facility (IRAF) procedure, including bias subtraction and flat-field correction. The aperture photometry was performed using the \texttt{APPHOT} package. The flux calibration was performed via differential photometry with field stars. The observations carried out on the same night were averaged. Finally we also used the archive data of the Catalina Sky Survey (CSS, \citealp{Drake09})\footnote{http://www.lpl.arizona.edu/css/}, which observed the source between 2005 and 2014 in V band with three dedicated telescopes. All the available photometric data points are illustrated in Fig.~\ref{fig:curva_ottico}.

\subsubsection{Spectroscopy}
We retrieved one spectrum from the Sloan Digital Sky Survey (SDSS) archive. It was observed by the SDSS/Barion Oscillation Spectroscopy Survey (BOSS, \citealp{Ahn14}) in 2012 (MJD 56090), in a lower activity state. Recent optical spectra were obtained with the Nordic Optical Telescope (NOT) and with the Telescopio Nazionale Galileo (TNG)\footnote{http://www.tng.iac.es/} on 2016 May 26 and 2016 May 30, respectively (see Table~\ref{tab:spettri}). In both cases the object was observed three times with an exposure time of 600 s. The spectra were reduced using standard IRAF tools. We first corrected for bias and flat-field, then we used a ThAr lamp to calibrate the NOT spectra and a HgNe lamp for the TNG spectra. Finally we flux calibrated both spectra using the standard star BD+332642. The final spectra have a dispersion of 2.20 \AA/px (NOT) and 2.68 \AA/px (TNG). To better compare the present-day spectrum and to study the source evolution in recent years, we also recovered spectra between years 2009 and 2013, obtained with the Andalucia Faint Object Spectrograph and Camera (ALFOSC) at NOT and the Intermediate Dispersion Spectrograph (IDS) at INT between 2009 and 2013. The observational details are shown in Table~\ref{tab:spettri}. Finally, we observed again the target on 2016 August 28 with the Asiago 1.22m telescope, with an exposure time of four hours. The spectrum was calibrated following the standard procedure already described, with a FeAr lamp for wavelength calibration and the standard star BD+332642 for flux calibration. The spectrum has a dispersion of 2.6 \AA/px. \par
We finally retrieved optical spectra observed by the Steward Observatory monitoring of \textit{Fermi LAT} sources \citep{Smith09}\footnote{http://james.as.arizona.edu/$\sim$psmith/Fermi/}. The radio quasar 3C 345 was observed 48 times between March 2009 and September 2012, covering a spectral range between 4000 and 7550\AA{} with a dispersion of 4\AA{}/px. The flux calibration was performed by scaling the spectra to match the V-band magnitude determined from the differential spectrophotometry on the same night. The spectral parameters measured on each spectrum are shown in Table~\ref{tab:spectral_parameters}. 
\begin{figure}[t!]
\centering
\includegraphics[width=\hsize]{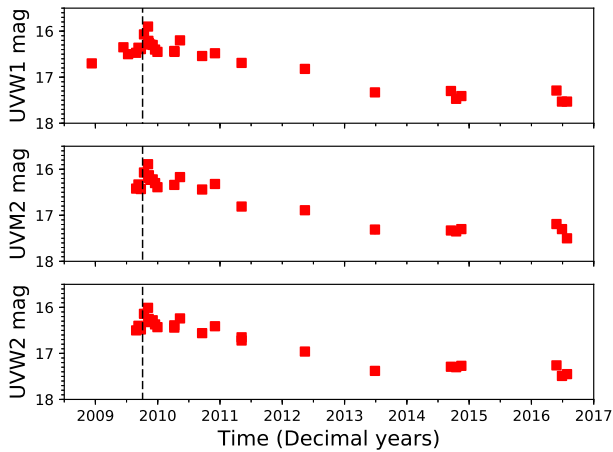} 
\caption{Light curve of 3C 345 in the UV bands measured by Swift/UVOT. From top to bottom, UVW1 magnitude, UVM2 magnitude, and UVW2 magnitude. The vertical dashed line indicates the 2009B flare.}
\label{fig:curva_uv}
\end{figure}
\subsection{UV and X-rays by Swift} 
\begin{figure}[t!]
\centering
\includegraphics[width=\hsize]{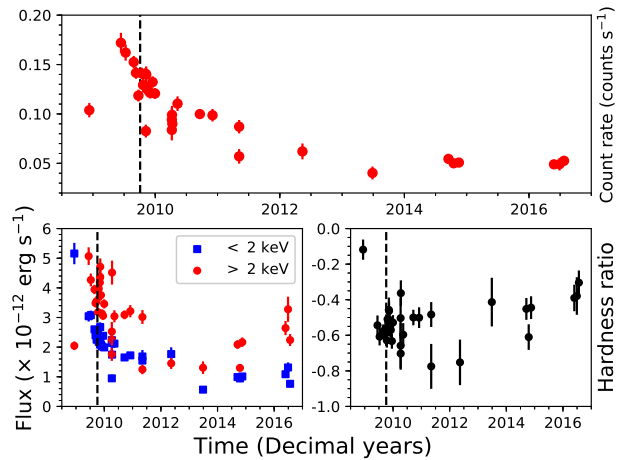} 
\caption{Top panel: Light curve of 3C 345 in X-rays, as derived from Swift/XRT data, in units of counts s$^{-1}$. The vertical dashed line indicates the 2009B flare. Bottom left panel: Light curve of the soft and hard X-ray flux, in erg s$^{-1}$. The threshold energy is fixed at 2 keV. Bottom right panel: Light curve of the hardness ratio, as defined in Eq.~\ref{eq:hardness}. The vertical dashed line indicates the 2009B flare.}
\label{fig:curva_X}
\end{figure}

We observed 3C 345 three times with the \textit{Swift} Satellite in target of opportunity mode. The first on 2016 April 25 with an exposure time 2.5 ks, again with 1.5 ks on 2016 May 27, and finally on 2016 June 23 with 2.5 ks. We also processed and analyzed all the archive data from Swift, to obtain historical X-ray and UV light curves of the source. The data reduction was performed using \texttt{HEASoft v.6.18}, with the calibration database updated on 2016 April 17. The data we analyzed were obtained with the X-Ray Telescope (XRT, \citealp{Burrows05}) and the UltraViolet Optical Telescope (UVOT, \citealp{Roming05}). Spectral counts from XRT were binned using at least 20 counts per bin, to apply the $\chi^2$ test. In two cases, too few counts were present to apply the $\chi^2$ statistic, therefore we used the maximum likelihood-based statistic for Poisson data \citep{Cash79}. The X-ray spectra analysis was performed using \texttt{XSPEC v12.9.0}, adopting a single power law model. The observation with UVOT was performed using all the six UVOT filters \citep{Poole08}. Measured magnitudes (VEGA flux zeropoints) were corrected for Galactic absorption following \citet{Cardelli89} and converted into physical units using the zero point magnitudes provided in the calibration database. The optical filter observations are presented in Fig.~\ref{fig:curva_ottico}, while the UV light curves are given in Fig.~\ref{fig:curva_uv}. The X-ray properties, instead, are summarized in Fig.~\ref{fig:curva_X} and in Table~\ref{tab:X_flux}. \par
\subsection{Gamma rays by Fermi LAT} 
To study the high-energy emission we used the public data archive of the Large Area Telescope (LAT) on board of the \textit{Fermi} $\gamma$-ray Space Telescope \citep{Atwood09, Ackermann12}. 
In order to take into account the full set of available observations, covering the most recent history of the source, we analyzed all the source-type events in the range between 0.1 and 500 GeV collected from the beginning of the \textit{Fermi} all-sky survey on 2008 August 4 and 2016 September 1. 
To reduce the data we used the LAT Science Tools \texttt{v10.0.5}, with the event selection evclass=128, evtype=3, and instrument response function (IRF) \texttt{P8R2\_SOURCE\_V6}, the Galactic diffuse background \texttt{gll\_iem\_v06}, and the isotropic background \texttt{iso\_P8R2\_SOURCE\_V6\_v06} to model the Galactic and isotropic diffuse emission. Contamination due to the $\gamma$-ray-bright Earth limb is avoided by excluding events with zenith angle $>$90$^\circ$. 

We followed a likelihood-based approach to obtain the photon fluxes at this band. 
Therefore to model the emission we considered a power law of the form, $F(E) \propto E^{-\Gamma}$, where $\Gamma$ is the photon index. 
In order to quantify the detections we accounted for the Test Statistic \citep[TS, ][]{Mattox96}, determined on the source in uniform time bins of 30 days, in such a way that any bin with TS $< 4$ was classified as an upper limit. \par
We first performed a fit of the whole time interval using unbinned analysis, modeling the target and those nearby sources already reported in the 3FGL catalog \citep{Ackermann15} within a region of interest (ROI) of 10$^\circ$. 
Within this ROI, 19 nearby sources were modeled along with 3C 345.
The latter is bright within the ROI, with a TS $=$ 1953, a photon index of 2.38$\pm$0.04, and a photon flux of (4.5$\pm$0.4)$\times$10$^{-8}$ ph cm$^{-2}$ s$^{-1}$. 
This result is consistent with the value reported in the 3FGL catalog. 
The $\gamma$-ray position is at coordinates R.A.$=$250.7348$^\circ$ and Dec.$=$39.816$^\circ$ (J2000). 
Residual TS maps were obtained to check whether there was any new emerging background source. 
Indeed, our analysis suggests the presence of two $\gamma$-ray sources not included in 3FGL, but relatively faint and distant from the target.
These two objects are located at R.A.$=$255.304$^\circ$, Dec.$=$40.027$^\circ$ and R.A.$=$248.878$^\circ$, Dec.=36.492$^\circ$.  
Repeating the analysis including these two background sources, and leaving both their normalization and spectral photon index parameters free to vary, we found a flux of  (4.3$\pm$0.4)$\times$10$^{-8}$ ph cm$^{-2}$ s$^{-1}$, with a TS $=$ 1890 and a spectral photon index of 2.36$\pm$0.04. \par
\begin{figure}[t!]
\centering
\includegraphics[width=\hsize]{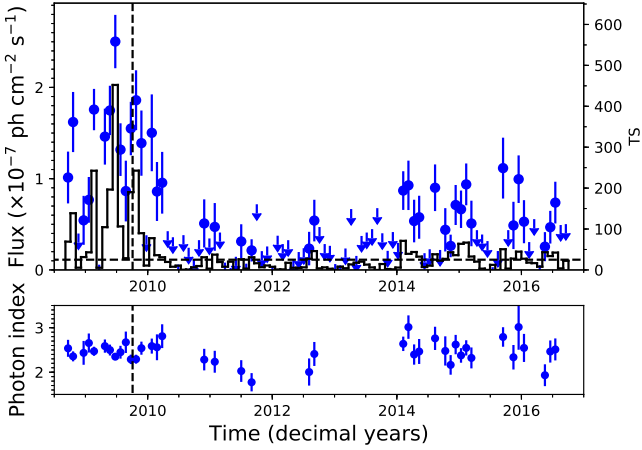} 
\caption{Vertical dashed line indicates the 2009B flare. \textbf{Top panel:} Monthly flux light curve of 3C 345 in $\gamma$ rays measured by \textit{Fermi LAT}. The bars indicate the TS level of each bin, shown on the right y-axis. The horizontal dashed line indicates the TS threshold of 25. \textbf{Bottom panel:} Monthly photon index light curve. Only photon indexes associated with points with TS$>$25 are shown.}
\label{fig:curva_gamma}
\end{figure}

We then calculated the monthly time bin light curve in the same approximately 8-year considered period. In each monthly time interval the spectral parameters of the known background sources were fixed to the 3FGL values. 
The same was done for all the background sources with a TS $<$ 1000. 
The results of this sampling are illustrated in Fig.~\ref{fig:curva_gamma}, along with the photon index measured in each bin. 
Our measurements are typically consistent with 3FGL catalog values, although with some discrepancies that may be accounted for by the use of Pass 8 IRF instead of the Pass 7 reprocessed IRF used for the 3FGL catalog, and by the presence of two nearby sources not included in the 3FGL. \par
To study the period of highest activity of the source, we also calculated the three-hour bin light curve around the peak registered on 2009 October 4, from three days before to three days after. 
We first performed a seven-day analysis, which provides a preliminary estimate of the target parameters and background sources. 
Then, in the light curve analysis, the spectral photon index of all the sources and the normalization of the diffuse background components were frozen. 
The spectral photon index value we used was -2.39. 
The binning we used for the light curve is three hours, and when TS $<$ 4 we considered the flux as an upper limit. 
The results are displayed in Fig.~\ref{fig:flare_gamma}. 

\section{Results}
\subsection{Light curves}
\begin{figure}[t!]
\centering
\includegraphics[width=\hsize]{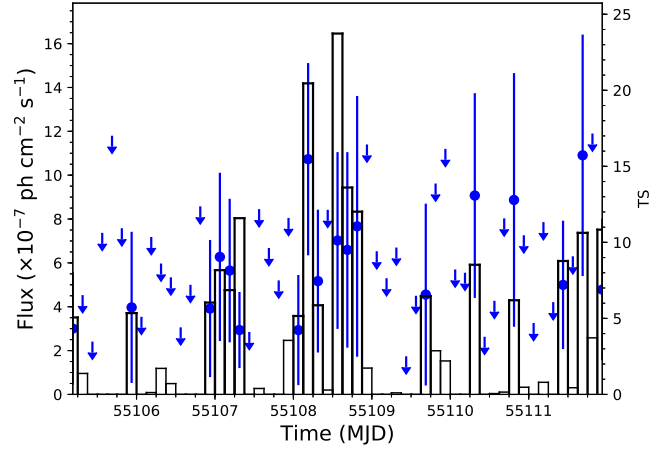} 
\caption{Flare from 2009 October 4 of 3C 345 in $\gamma$ rays. Each bin corresponds to three hours. The bars indicate the TS level of each bin, shown on the right y-axis. When the TS was $<$4, we considered the flux as an upper limit.}
\label{fig:flare_gamma}
\end{figure}
\begin{figure}[t!]
\centering
\includegraphics[width=\hsize]{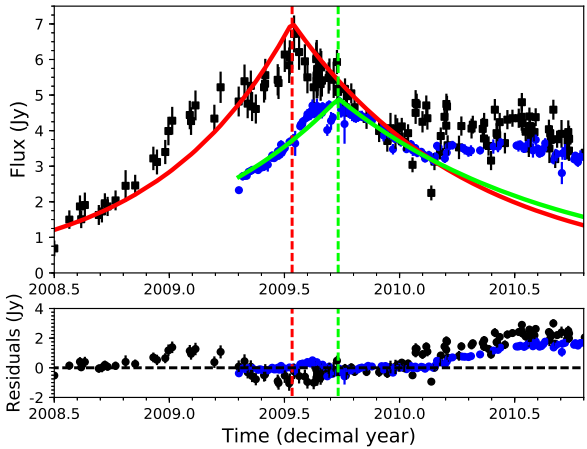} 
\caption{Exponential models of the radio flare 2009A. \textbf{Top panel:} Black squares indicate measurements at 37 GHz, the blue circles indicate 15 GHz data, after continuum subtraction. The red solid line is the exponential model of Eq.~\ref{eq:exp}, and describes the 37 GHz flare, while the red dashed vertical line indicates the event peak. The green solid line models the 15 GHz flare, and the green dashed vertical line indicates the flare peak. \textbf{Bottom panel:} Residuals of the data after the subtraction of the exponential model. Colors are as in the top panel.}
\label{fig:flare}
\end{figure}

The data analysis revealed the presence of at least two prominent flares in 2009, which propagated from high energies to lower frequencies, and of a very low state in 2016 in all the frequencies. In the following we will investigate the light curves at different frequencies.\\
\textbf{Radio band:} The radio light curves at 37 and 15 GHz are shown in Fig.~\ref{fig:curva_radio}. To properly model the flaring event of 2009A, we adopted a technique to decompose variations into exponential flare peaks following \citet{Valtaoja99}. We subtracted from each light curve the continuum level, identified as the minimum flux measured in the source during the examined period, and we fitted it with a curve in the exponential form
\begin{equation}
\Delta S(t) = \begin{cases}  \Delta S_{max} e^{(t - t_{peak})/\tau}\\\Delta S_{max} e^{(t_{peak} - t)/1.3\tau} \end{cases} \; ,
\label{eq:exp}
\end{equation}
where $\Delta S_{max}$ is the maximum amplitude of the flare, $t_{peak}$ is the peak flux epoch, and $\tau$ represents the flux rise timescale. 
The model was applied to the 37 GHz curve and the 15 GHz curve. 
The results are shown in the top panel of Fig.~\ref{fig:flare}. 
At 37 GHz, the peak flux was reached on 2009 July 05, while the peak of the 15 GHz flare is reached on 2009 September 24. 
The rise timescale of this event was 0.58 years at 37 GHz, and 0.72 years at 15 GHz, corresponding to 0.36 and 0.45 years in the source restframe ($\tau_{rest} = \tau/(1 + z)$, where $\tau_{rest}$ is the timescale in the source frame). 
The residuals, shown in the bottom panel of Fig.~\ref{fig:flare}, indicate that after the event there is a significant flux increase, which cannot be modeled with an exponential profile.
This reached a peak at the beginning of 2010, and that later decayed into the low state. 
We interpret this as a second outburst of the source, which is associated with analog events at different frequencies occurring at the end of 2009. 
Therefore, from now on, we will refer to these two events as 2009A and 2009B. 
It is worth noting that before the 2009A event, the 37 GHz light curve shows another flux increase. 
However, our data do not allow us to fully constrain the position of its peak. \par
The light curve showed another outburst on 2015 December 25 at 37 GHz, and 2016 March 26 at 15 GHz. 
The aftermath of this event is well described by an exponential profile, with characteristic timescales of $\tau =0.57$ years at 37 GHz, and 0.28 years at 15 GHz (0.36 and 0.18 years in source frame). 
Unfortunately the only multiwavelength data available were the continuous $\gamma$-ray monitoring of the \textit{Fermi LAT}. \par

\textbf{Optical:} The curve is shown in Fig.~\ref{fig:curva_ottico} for five optical filters, U, B, V, R, and I. 
In 2009 the strong activity of the source is visible in all filters. 
The flares cannot be modeled with Eq.~\ref{eq:exp}, probably because of the small number of data points. 
A careful inspection of the light curves seems to indicate the presence of both the 2009 outbursts. 
The I filter in particular clearly shows two peaks in the data, the first on 2009 May 15, and the second on 2009 September 12. 
A similar shape is visible also in the other filters. 
It is worth noting that there might be a third outburst, visible only in the R and I filters, with a peak measured on 2010 April 07. 
It is not clear whether this is associated with counterparts at other frequencies. 
However, the $\gamma$-ray monthly light curve shows a slight increase in the same time bin, with respect to the previous and following months. \par

\textbf{Ultraviolet:} Figure~\ref{fig:curva_uv} shows the behavior of the three UV filters of Swift/UVOT. 
The brightest magnitude reached in our time range was measured in 2009 (see Table~\ref{tab:mag_uv}), while the faintest magnitudes were measured in June 2016. 
The UV data reveal the presence of the 2009B flare only, probably because the source was not monitored during May 2009. 
No other event is visible in the Swift/UVOT data. \par

\textbf{X-rays:} In this band we also evaluated the hardness ratio, defined as
\begin{equation}
HR = \frac{H - S}{H + S} \; ,
\label{eq:hardness}
\end{equation}
where $H$ and $S$ represent the flux above and below 2 keV, respectively. The light curve, both in count rates and in physical units, along with hardness ratio and hard and soft X-rays flux light curves, is shown in Fig.~\ref{fig:curva_X}. \par
As in UV and optical, X-rays show a clear decreasing trend in flux since 2009 to the present day. 
The highest flux of X-rays was registered on 2009 May 11, and therefore it is likely to be associated with 2009A event. 
The luminosity then seems to decrease, but it shows a slight increase in correspondence with the optical/UV flare 2009B. 
The hardness ratio instead showed a high flux of hard photons in 2008 November 09, when no particular activity was measured at different wavelengths.
During the other periods, the hardness ratio was relatively stable, showing a slight increase in the low state. 
The lowest flux in X-ray band was observed on 2013 May 26, when neither in optical nor in radio was a particularly low state observed.
Finally, the X-ray photon index, shown in Table~\ref{tab:X_flux}, does not vary significantly during the examined period, remaining approximatively constant within the error bars. \par

\textbf{Gamma rays:} The resulting light curve is shown in Fig.~\ref{fig:curva_gamma}. 
Our results are consistent with those of 3FGL in the overlapping time range. 
The 2009 flares are both visible, with a noticeable increase of the high-energy flux. 
A high flux was reached during flare 2009B, on modified Julian date (MJD) 55108.1875 (2009 October 04, 04h30m00s), with a value of (1.07$\pm$0.44)$\times10^{-6}$ photons s$^{-1}$ cm$^{-2}$ and a TS$\sim$20. 
As shown in Fig.~\ref{fig:flare_gamma}, a substantial amount of energy was emitted during the 2009B outburst. 
The activity later decreased, although a comparably high photon flux, (1.09$\pm$0.55)$\times10^{-6}$ photons s$^{-1}$ cm$^{-2}$, was measured also on 2009 October 7 with a lower significance (TS$\sim$10). \par
An increase in $\gamma$-ray activity of the source was measured from the beginning of 2014 to the end of the examined period. 
The highest flux of this event is measured in January 2016, with a flux (1.15$\pm$0.24)$\times10^{-7}$ photons s$^{-1}$ cm$^{-2}$ and a TS$\sim$58. 
This might be associated with the increased radio emission observed during the same period. 
However, no other data are available to clearly establish how they are related. \par
The photon index, shown in the bottom panel of Fig.~\ref{fig:curva_gamma} and defined as F$_E \propto E^{-\Gamma}$, is roughly constant in the monthly bins which are associated with a high state, remaining approximately close to $\Gamma \sim$2.5. 
The measurement errors are too large to allow us to make any claim for photon index variability in the two emission states. \par
To estimate the timescale of rapid $\gamma$-ray variability, we calculated the doubling timescale during the 2009B event using the intraday light curve (Fig.~\ref{fig:flare_gamma}), as
\begin{equation}
\tau_d = \frac{(t - t_0) \log 2 }{\log F(t) - \log F(t_0)} \; ,
\label{eq:doubling}
\end{equation}
where $\tau_d$ is the doubling timescale, $(t - t_0)$ is the time interval between two flux measurements, and $F(t)$, $F(t_0)$ are the flux values. At the beginning of the flare, the flux changed from 2.93$\times10^{-7}$ ph s$^{-1}$ cm$^{-2}$ to 1.07$\times10^{-6}$ ph s$^{-1}$ cm$^{-2}$. According to Eq.~\ref{eq:doubling}, $\tau \sim$1.6 hours, that is approximately one hour in the source frame. 
Taking into account the uncertainties on the flux, we can constrain the doubling time between 0.6 and 13.6 hours, corresponding to $\sim$0.4 and 8.5 hours, respectively, in the source frame. 

\subsection{Cross-correlation}
In order to study the behavior of the source at different frequencies, we performed a cross-correlation analysis on the light curves. Since we had to deal with unevenly sampled data, we decided to use the discrete correlation function (DCF, \citealp{Edelson88}). The DCF is evaluated as
\begin{equation}
DCF_{ij} = \frac{(x_i - \overline{x})(y_j - \overline{y})}{\sigma_x\sigma_y} \; ,
\end{equation}
where $x_i$ and $y_j$ are the observed fluxes at times $t_i$ and $t_j$, $\overline{x}$, $\overline{y}$ are the mean fluxes in the two observed bands, and $\sigma_x$, $\sigma_y$ are the standard deviations of the light curves. 
We tested the chance probability of the correlations by generating 5000 light curves with underlying red noise properties and a distribution resembling those of the observed data. 
These surrogate light curves were then cross-correlated in a manner similar to those performed with the observed data, thus yielding a distribution at every time lag bin. 
From these distributions, the 1, 2 and 3$\sigma$ significance interval was obtained.
A detailed description of the procedure we used is given by \citet{Ramakrishnan15}. \par
\begin{figure}[t!]
\centering
\includegraphics[width=\hsize]{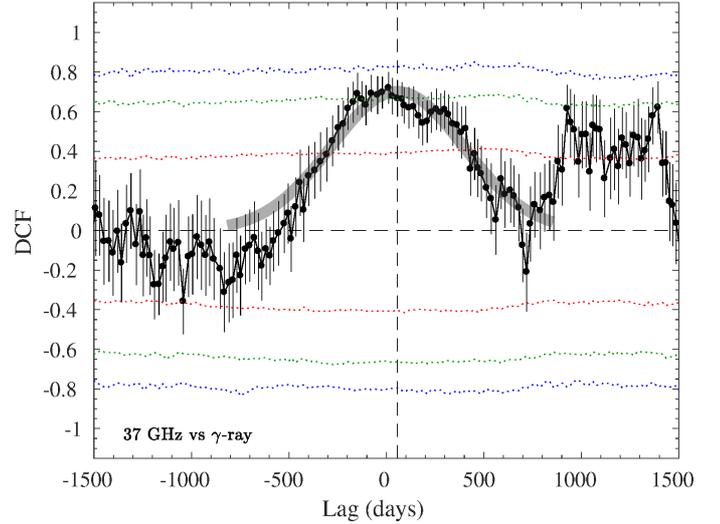} 
\caption{Discrete cross-correlation function between $\gamma$ rays and 37 GHz flux. The red, green, and blue dotted lines indicate a significance level of 1, 2, and 3$\sigma$, respectively. The horizontal dashed line indicates the zero level of the DCF. The vertical dashed line indicates the peak position of a Gaussian fit to the correlation curve at 56 days. The Gaussian fit we performed on the DCF is indicated by the shaded gray line.} 
\label{fig:cross_V_radio}
\end{figure}

The correlation between the radio and $\gamma$-ray light curve in 3C 345 is shown in Fig.~\ref{fig:cross_V_radio}.
The data we analyzed are the 37 GHz radio flux, illustrated as the black points of Fig.~\ref{fig:curva_radio}, and the $\gamma$-ray fluxes shown in Fig.~\ref{fig:curva_gamma}.
In our analysis, the peak from the Gaussian fit to the correlation curve indicates that the radio lags the $\gamma$ rays by 56 days at a significance level of 2$\sigma$.
The same investigation was already carried out in \citet{Ramakrishnan15} on a different time interval. With a weekly binning for $\gamma$ rays, they found two significant lags, the first of -40 days, meaning that the radio led the $\gamma$-ray curve, and a second one with 80 days. Using instead a monthly binning for the $\gamma$-ray curve they found only one significant lag at 30 days, with $\gamma$ rays preceding the radio emission. 
The latter result is in agreement with ours, since radio flux seems indeed to respond to optical and high energy, particularly in correspondence to flaring events. \par
We performed the cross-correlation using optical data, but we found instead no significant correlations. 
A similar result was found between $\gamma$ rays and V magnitude. \par
It is however worth noting that, in the light curve, there is clear evidence that the optical and $\gamma$-ray flares are nearly simultaneous. In particular, in the event 2009B, the optical and high-energy emissions reached a peak on the same day. \par
\begin{figure}[t!]
\centering
\includegraphics[width=\hsize]{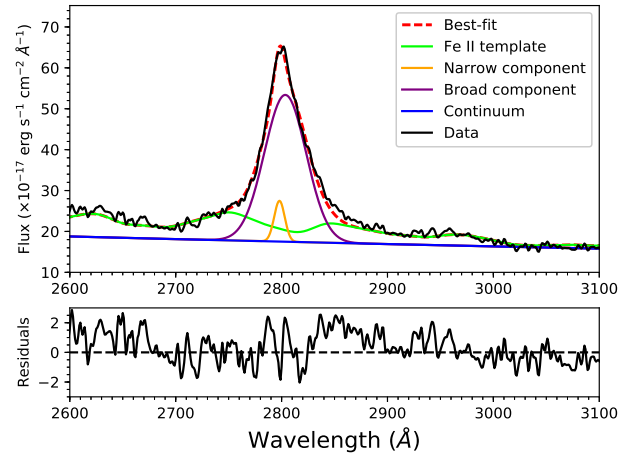}
\caption{Top panel: Fit of the Mg II region in the 2016 May 30 TNG spectrum. The black solid line is the observed flux, the blue solid line indicates the continuum power law (slope $\alpha_\lambda = -1.02\pm 0.08$), the green solid line represents the Fe II multiplets, the orange solid line is the narrow component of Mg II, the purple solid line indicates the Mg II broad component, while the red dashed line represents the best-fit model, obtained by adding all of the previous components. Bottom panel: Residuals between the data and the model, in units of $10^{-17}$ \ergs\ \AA$^{-1}$.}
\label{fig:fe_subtraction}
\end{figure}
\subsection{Black hole mass}
We used the optical spectra to derive the black hole mass and the Eddington ratio. The H$\beta$ line is often used to perform this kind of estimate. However, in our case it lies too close to the atmospheric telluric absorptions. Reconstructing the line profile so close to this feature might introduce unquantifiable errors in the measurements. Therefore, we decided to use the Mg II line, which is present in all of our spectra. \par
We corrected all of our spectra for Galactic absorption, using N$_H = 1.06\times10^{20}$ cm$^{-2}$ \citep{Kalberla05}, which corresponds to A(V) = 0.056, and redshift. We then subtracted the continuum around the line reproducing it with a power law. To estimate the continuum spectral index, we repeated the fitting procedure several times by changing the fixed continuum regions, and considered the median value and its associated standard deviation (see Table~\ref{tab:spectral_parameters}). A crucial step is the subtraction of the Fe II multiplets surrounding Mg II. As a template model, we used that produced by \citet{Bruhweiler08} with density $\log$N$_H$ = 12 cm$^{-3}$, microturbulence of 20 \kms, and ionizing flux 10$^{20.5}$ cm$^{-2}$ s$^{-1}$\footnote{http://iacs.cua.edu/personnel/personal-verner-feii.cfm}. Following \citet{Marziani13a, Marziani13b} the template was convolved with Gaussians of different widths to broaden the lines, and we subtracted the best-fit template from the spectrum with the highest signal-to-noise (S/N) ratio in the $\lambda$3000 continuum, which is the TNG spectrum of 2016 May 27 (S/N $\sim$ 19). We later subtracted the same template from the other spectra as well, rescaling its flux in order to obtain an optimal result. \par
To decompose the line profile we used two Gaussians. One was used to reproduce the broad component, while the second was used for the narrow component. We fixed the FWHM of the narrow component to be the same as the line of [Ne III]$\lambda$3869, which is visible in the TNG spectrum, since the narrow-line region is not expected to vary significantly during the time interval of our data. This width is $\sim$855 \kms. Not accounting for the very broad component of the Mg II line might affect the black hole mass measurement \citep{Marziani13a}. However, our spectra do not all have a S/N high enough to disentangle more than two components, so we preferred to avoid an overfitting of the line. To derive the black hole mass we followed \citet{Shen12}, 
\begin{equation}
\log\left(\frac{M_{BH}}{M_\odot}\right) = a + b\log\left(\frac{L_{\textrm{line}}}{10^{44}\;\textrm{erg s}^{-1}}\right) + c\log{\left(\frac{\rm FWHM_b}{\textrm{km s}^{-1}}\right)}\; ,
\label{eq:massa_fwhm}
\end{equation}
where L$_{line}$ is the luminosity of the Mg II line, FWHM$_b$ is the broad component FWHM corrected for instrumental resolution, and $a$, $b$, $c$ are three empirically determined constants which are 3.979, 0.698 and 1.382, respectively. The line luminosity is evaluated on the Gaussian fit. We derived a black hole mass estimate in each spectrum. To evaluate the error induced by the fitting procedure, we repeated the fitting 100 times by introducing a Gaussian noise proportional to that measured in the $\lambda$3000 continuum. The median value is 6.17$\times$10$^8$ M$_\odot$ ($\log{(M/M_\odot)} =8.79$), with a standard deviation among the different measurements of 0.17 dex. \par
The black hole mass of 3C 345 has been calculated several times in the literature, with values spanning between $10^6$ M$_\odot$ \citep{Zhang98} to the much higher estimate provided by \citet{Lobanov05} of two black holes of 7$\times$10$^8$ M$_\odot$ each. Our estimate is however in good agreement with the upper limit provided by \citet{Liu15} and derived through the $\gamma$-ray variability timescale  ($\log{(M/M_\odot)} = $ 9.10), and with those provided by \citet{Xie05} and \citet{Ghisellini14} ($\log{(M/M_\odot)} = $ 8.41, and 9.03, respectively).
\begin{figure}[t!]
\centering
\includegraphics[width=\hsize]{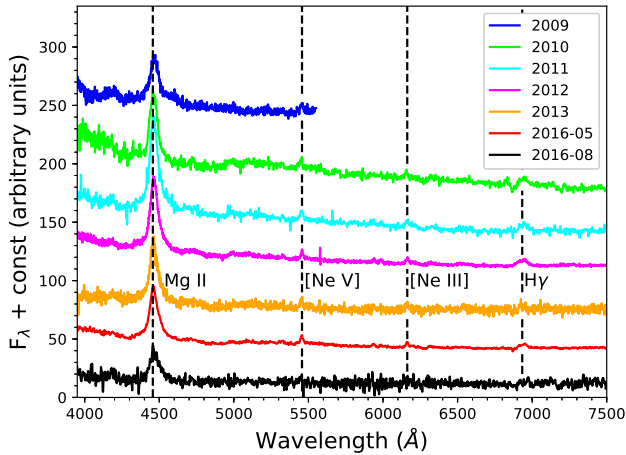}
\caption{Optical spectra of 3C 345 obtained between 2009 and 2016 with NOT, INT, TNG and Asiago telescopes. The vertical dashed lines indicate the most prominent emission lines. Flux in arbitrary units.}
\label{fig:spettro}
\end{figure}
\subsection{Spectral lines and continuum variability}

\begin{figure}
\centering
\includegraphics[width=\hsize]{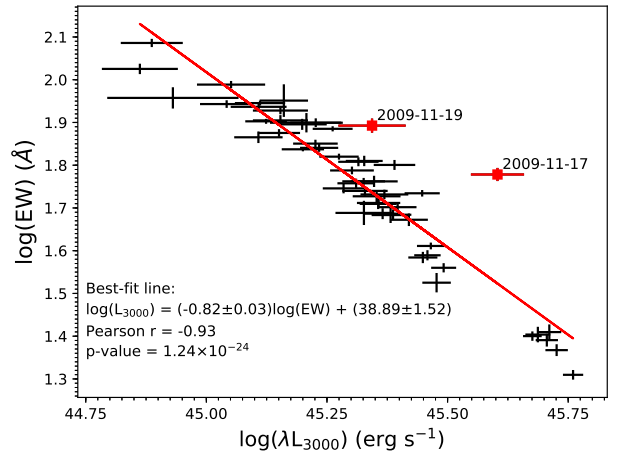} 
\caption{Anticorrelation between continuum flux at 3000\AA{} and equivalent width of Mg II $\lambda$2798 emission line. The two red points indicate the data for spectra observed on 2009 November 17 and 2009 November 19, immediately following the 2009B flare.}
\label{fig:corr_ew_cont}
\end{figure}
\begin{figure}
\centering
\includegraphics[width=\hsize]{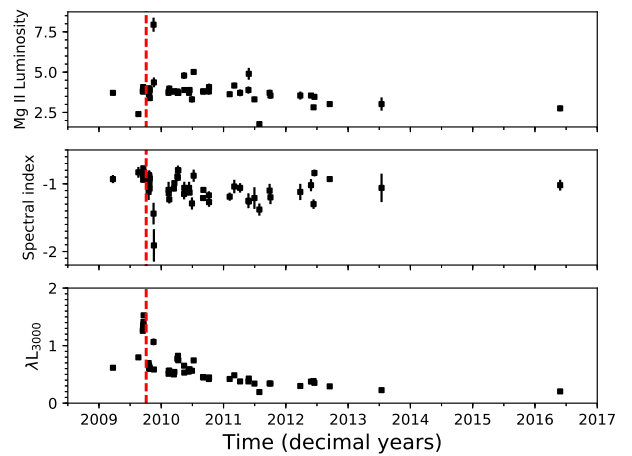} 
\caption{\textbf{Top panel:} Light curve of the total Mg II $\lambda$2798 line flux. The flux is in units of 10$^{42}$ erg s$^{-1}$. \textbf{Middle panel:} Time evolution of optical spectral index. \textbf{Bottom panel:} Light curve of the 3000\AA{} continuum luminosity, in units of 10$^{41}$ erg s$^{-1}$. The red vertical dashed line in each plot indicates the 2009B flare.}
\label{fig:mg_lightcurve}
\end{figure}

Exploiting the Steward observatory archival spectra and our new observations (see Fig.~\ref{fig:spettro}), we measured the equivalent width (EW) of the Mg II line, and compared it to the $\lambda$3000 continuum luminosity. 
The line EW was measured on the redshift-corrected spectra after subtracting only the Fe II multiplets. 
We estimated it by evaluating the line profile and the continuum level 100 times for each spectrum, adding every time a different Gaussian noise proportional to that in the $\lambda$3000 continuum.
The measurement we adopt is the median of the obtained distribution, while its standard deviation is the error associated to our estimate. 
In our measurements, the EW of an emission line is defined as positive. 
The EW and the continuum luminosity are anticorrelated (e.g. \citealp{Foschini12b}). 
A higher continuum flux induced by the rapid variability of the jet indeed reduces the EW of the lines. 
This result is clearly visible in the Mg II line, and it is shown in Fig.~\ref{fig:corr_ew_cont}. \par
The Pearson coefficient we measured is -0.93. The p-value, that is the probability of the correlation being generated by chance, is 1.6$\times$10$^{-24}$ when all the points are included. 
However, two points, those derived after the 2009B outburst, seem to show a different behavior (see Sect.~\ref{sec:discussion}). 
When they are excluded, the Pearson coefficient increases to -0.96, with a p-value of 3.7$\times10^{-30}$, as reported in Fig.~\ref{fig:corr_ew_cont}, indicating an even stronger correlation.  \par
We derived the spectral slope with the \texttt{nfit1d} task of IRAF.
The results, along with the light curve of Mg II and $\lambda$3000 flux, are shown in Fig.~\ref{fig:mg_lightcurve} and in Table~\ref{tab:spectral_parameters}.
After the flare, the optical spectrum became much steeper, going from a slope of $\alpha_\lambda \sim -1$ (F$_\lambda \propto \lambda^{\alpha_\lambda}$) to $\alpha \sim -2.17$, forty-six days after the flare ($\sim$29 days in the source frame). 
At the same time, the Mg II luminosity doubled, reaching a value of $\sim$7.5$\times$10$^{42}$ \ergs. 
While the $\gamma$ rays and the continuum luminosity peak are reached on 2009 October 4, the Mg II line responded much later. 
Our spectra in that period were observed every night between 2009 October 19 and 2009 October 26, and later on 2009 November 17. 
The eight consecutive spectra show an increasing trend in the 3000\AA{} continuum luminosity, in agreement with the decreasing optical magnitude measured for the source. 
In contrast, on 2009 November 17, the Mg II flux was twice as large relative to neighboring observations, suggesting that on that day, the increase in the flux of ionizing photons had already reached the region where Mg II is formed (i.e., the broad-line region, BLR).
Assuming that the outburst event was produced on 2009 October 4, we can constrain the distance between the region of production of the $\gamma$-ray photons (possibly, but not necessarily, close to the central engine) and the BLR between 22 and 44 light-days (14 and 28 days in the source frame, respectively). 
The increase in Mg II flux indeed occurred at some point between 2009 October 26, the last spectrum observed in which the line flux was not affected yet by the outburst, and 2009 November 17, the first spectrum showing an enhanced line luminosity. 
We measured directly the BLR radius from the low-state spectrum of 2016, following the relation developed by \citet{Greene10},
\begin{equation}
\log\left(\frac{R_{\mathrm{BLR}}}{10 \textrm{ ld}}\right) = 0.85 + 0.53\log\left[\frac{L(\textrm{H}\beta)}{10^{43} \textrm{erg s}^{-1}}\right] \, .
\end{equation}
The luminosity of H$\beta$ was calculated by fitting it with two Gaussians, as done for Mg II. 
As mentioned before, there might be a nonnegligible error in the measurement due to the vicinity of the telluric absorption, therefore this estimate is only a rough approximation. 
The result is R$_{BLR} \simeq 46$ ld, which is a factor $\sim$2 larger than the estimate obtained from the Mg II reverberation. \par
\subsection{Spectral energy distribution}
\begin{figure}
\centering
\includegraphics[width=\hsize]{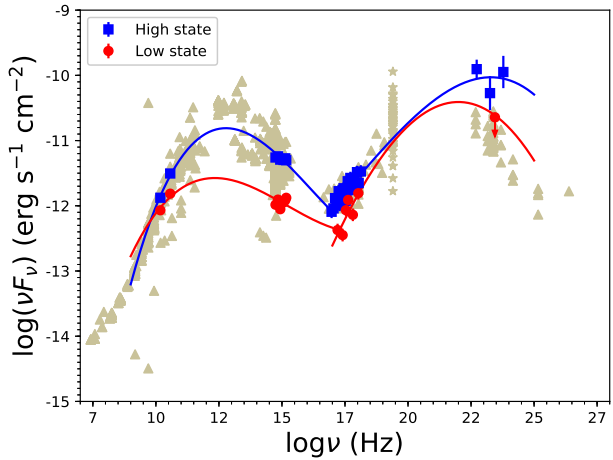} 
\caption{Simultaneous spectral energy distributions of 3C 345. The blue points indicate the high state of 2009 October 4 (2009B flare), while the red points indicate the low state 2016 June 26. The blue and red solid lines are the best fit with a polynomial curve for high and low state, respectively. The light gray triangles are retrieved from the public archive of the ASI SSDC SED builder tool, the light gray stars are derived from the 70-months BAT all-sky hard X-ray survey \citep{Baumgartner13}.}
\label{fig:SED}
\end{figure}

The SEDs of both the high state of 2009 October 4 and of the low state of 2016 June 26 of 3C 345 are shown in Fig.~\ref{fig:SED}, along with other literature data derived from the Space Science Data Center of Agenzia Spaziale Italiana (ASI SSDC) SED builder archive\footnote{http://tools.asdc.asi.it/SED/} and the 70-month BAT all-sky hard X-ray survey \citep{Baumgartner13}.
We fitted separately the synchrotron and inverse Compton (IC) humps using a polynomial function of order three in both states. We assumed that the X-ray emission is part of the IC hump. Because of the small number of points, the order three polynomial diverges in the high-energy hump of the low state, not providing a physically reasonable result. Therefore, in that case only, we forced the polynomial to have a negative curvature and to fit the $\gamma$-ray upper limit. We then determined the peak frequency of the two humps in both cases. In the high state, $\log\nu_s \approx 12.78$, and $\log\nu_{IC} \approx 23.28$. In the low state, $\log\nu_s \approx 12.33$ and $\log\nu_{IC} \approx 22.00$. The SED appears to move toward higher frequencies during the flaring activity. \par
We calculated the physical properties of the source in the two different states, assuming a leptonic model for the jet. The main mechanism to produce the IC is likely to be external Compton, in which the BLR provides the seed photons. We assumed a bulk Lorentz factor of the jet of $\Gamma\sim10$, as derived by \citet{Lister13} from kinematical measurements of jet knots. In the following we also assumed that the peak of the external photon field $\nu_{ext}$ is the Ly$\alpha$ $\lambda$1216 line frequency, and that the difference in Doppler factor between the photon field and the electrons is $\Delta\delta \sim \Gamma$. The relativistic factor of the dominant electron population, as shown in \citet{Boettcher}, is then given by
\begin{equation}
\gamma_b = \sqrt{\frac{3\nu_{IC} (1 + z)}{4\nu_{ext}\Gamma \Delta\delta}} \; .
\end{equation}
The resulting factor moves from $\gamma_b \sim 963$ in the high state to $\gamma_b \sim 220$ in the low state. Once $\gamma_b$ is known, we estimated the Larmor frequency $\nu_L$ according to
\begin{equation}
\nu_L = \frac{3\nu_s (1 + z)}{4 \gamma_b^2 \delta} \; ,
\label{eq:larmor}
\end{equation}
where the Doppler factor $\delta$ is calculated using $\theta = 7^\circ$ \citep{Lobanov05}. The Larmor frequency enables us to derive the magnetic field of the emitting region as 
\begin{equation}
B = \frac{2\pi m_e c \nu_L}{e} \; ,
\label{eq:campomag}
\end{equation}
where $m_e$ and $e$ are the mass and the charge of the electron, respectively. The magnetic field thus is 0.27 G in the high state, and 1.91 $G$ in the low state. The latter result is in agreement with observed magnetic fields around 1 G for FSRQs \citep{Ghisellini15}. Moreover, the high state measurement is consistent with the 0.36 G observed by \citet{Matveyenko13} for 3C 345 outbursts of 1998-1999, and with the results found for other FSRQs during outbursts \citep{Pacciani14}. \par
It is worth noting that the SED seems to reveal an interesting feature in the optical/UV region. In the high state the emission in this region does not show any particular feature, while a high frequency flux excess appears in the low state. This emission is likely to originate from the accretion disk, which in the high state is not visible because of the overwhelming synchrotron component. This effect is essentially identical to the EW decrease in the emission lines of optical spectra. The disk emission seems instead not to be present in the X-ray spectrum of the low state. However, the statistic is too sparse to understand whether its absence is real. \par
\section{Discussion}
\label{sec:discussion}
During the period we examined, which spans eight years from August 2008 to August 2016, our light curves reveal three main flaring events.
Two of them began, in optical and $\gamma$ rays, in 2009 (2009A and 2009B in May and October, respectively). 
The third one took place at the end of 2015, but it was covered by our data only in radio and in $\gamma$ rays. 
\citet{Schinzel12} observed with VLBI during the 2009 activity periods the ejection of three plasma condensations, each one corresponding to an increased optical and $\gamma$-ray activity. 
Our results are consistent with theirs. 
The 37 GHz light curve reveals the presence of two main peaks, and the flux excess observed in our radio data before the 2009A event might be associated with the first plasma ejection (see bottom panel of Fig.~\ref{fig:flare}). 
The 15 GHz light curve displays the same behavior, following with a 0.2-year delay in the source frame the same pattern of the 37 GHz curve. 
This outburst activity was observed in UV and X-rays, as well as in optical. 
As shown in Fig.~\ref{fig:cross_V_radio}, and found by \citet{Ramakrishnan15} as well, there is a correlation between the $\gamma$-ray flux and the radio, with the former leading the latter by 56 days. \par
Our results are well described by a widely known model for FSRQ flares. 
The high states are associated with the ejection of a new superluminal component from the jet-base, which moves downstream along the jet.
After the ejection, the superluminal plasma is initially optically thick for radio, and it carries enough energy to produce $\gamma$ rays.
When they move away from the central engine, the ejecta become optically thin for synchrotron emission, causing the radio luminosity peak, but they are not able to emit high-energy photons anymore. \par
The strong anticorrelation between the continuum and the EW of the optical lines can also be connected with the $\gamma$-ray activity of the source. 
If the accretion disk and the jet are correlated \citep[e.g.][]{Ghirlanda11}, the ejection of a new blob will correspond to an increased luminosity both in $\gamma$ rays and in optical. 
The increased activity in the accretion disk will likely lead to a bluer and stronger continuum. 
Therefore, the Mg II line will respond decreasing its EW because of the increased continuum level, and later with an increased flux due to a reverberation effect \citep{Blandford82}.
This behavior is similar to that observed in BL Lacertae, where emission lines are visible on some occasions during low states of jet activity \citep{Vermeulen95}. \par
However, the difference in the EW after the flare might not be due to only a different jet activity, but instead to a change in the accretion disk. 
As shown in Figs.~\ref{fig:corr_ew_cont} and \ref{fig:mg_lightcurve}, both of the data points observed immediately after the flare seem indeed to show a different behavior with respect to the other points.
Their EWs indeed correspond to those expected for a less prominent continuum. 
A possibility is that the disk contribution to the optical spectrum is in this case very different than in the other spectra, because of an enhanced disk luminosity. \par
\begin{figure}
\centering
\includegraphics[width=\hsize]{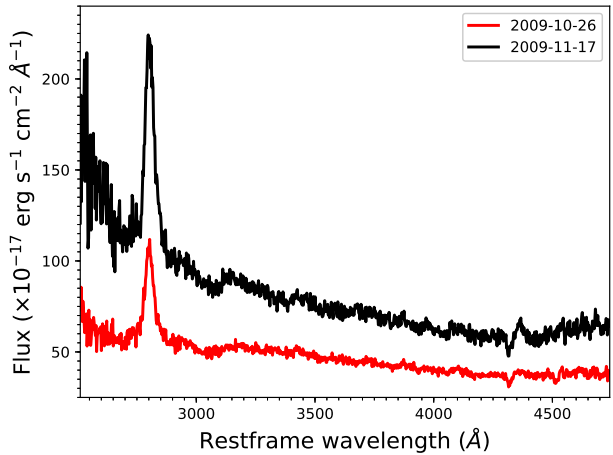} 
\caption{Optical spectra in restframe wavelength observed in two different days. The prominent emission line is Mg II $\lambda$2798. The red line indicates the spectrum observed on 2009-10-26, the black line indicates the spectrum obtained on 2009 November 17. Both spectra come from the public archive of Steward observatory.}
\label{fig:spind}
\end{figure}

The optical spectra we analyzed might also provide information on the position of the dissipation region in the 2009B outburst.
The region where $\gamma$-ray photons are produced has been the subject of a large debate in the last years. 
Some authors believe that the detection of very high-energy photons during outbursts in FSRQs is a sign that the dissipation region lies outside the BLR \citep{Donea03, Aleksic11a, Aleksic11b, Tavecchio13, Coogan16}. 
In our case, both optical continuum and $\gamma$ rays reached the maximum flux on the same day, 2009 October 4, as confirmed by the intra-day light curve in $\gamma$ rays (Fig.~\ref{fig:flare_gamma}) and the optical light curves. 
This confirms that the emission region at these frequencies is co-spatial. 
Optical spectra provide further details by means of Mg II reverberation, continuum luminosity, and spectral slope. 
The line luminosity indeed remained essentially unperturbed for 22 days after the flare ($\sim$14 days in the source frame, corresponding to a light-travel $\sim$0.01 pc), possibly a sign that the ionizing continuum had not yet reached the region where the line is produced (i.e., the BLR). 
In 22 more days, instead, the Mg II luminosity went from 2.2$\times10^{43}$ to 4.0$\times10^{43}$ erg s$^{-1}$ (see Fig.~\ref{fig:spind}).
The most straightforward explanation for this remarkable flux doubling is that at some point between the two observations the enhanced ionizing continuum had reached the BLR, producing new Mg II ions and increasing the line luminosity.
Such lag between the $\gamma$-ray emission and the Mg II luminosity indicates that the production of high-energy photons occurred at a distance $d \lesssim 0.02$ pc from the BLR, fairly close to the central black hole. 
This is in agreement with both observations and theoretical models based on the rapid variability observed in FSRQs and the observations of breaks in their high-energy spectra \citep{Finke10, Tavecchio10, Lei15, Liao15a}. 
The doubling timescale we evaluated using Eq.~\ref{eq:doubling} for 2009B,  approximately one hour in the source restframe, is very short. 
Such a fast variability was observed only in a few objects \citep{Foschini11a, Foschini13, Vovk13, Liao15a} and, in agreement with the Mg II reverberation, it might indicate that the dissipation region is located within the BLR. 
However, as suggested by \citet{Foschini11a}, it is possible that, even in the same source, the location of $\gamma$-ray production changes for different outbursts. \par
The SED of 3C 345 can be efficiently described by a one-zone leptonic model with an external Compton contribution.
The source during high and low state shows the typical behavior of FSRQs, with both the IC and synchrotron peaks moving from higher to lower frequencies, respectively. 
The decrease of $\sim$1 G of the magnetic field intensity during the high state is a consequence of the different $\gamma_b$ of the electrons.
An increase in $\gamma_b$ leads to a lower Larmor frequency in Eq.~\ref{eq:larmor}, and hence to a lower magnetic field according to Eq.~\ref{eq:campomag}, as we indeed observe. \par
The strong change in the magnetic field intensity might be connected with another important aspect inferred from the optical spectra.
As clearly visible in Fig.~\ref{fig:spind}, 44 days after the flare the spectrum showed a significantly steeper continuum, going from an index of $-0.85\pm0.03$ on 2009 October 26 to $-1.50\pm0.05$ on 2009 November 17, with a further steepening to $-2.17\pm0.09$ on 2009 November s19.  
This behavior is probably not associated with the Mg II luminosity increase, even if they occur almost at the same time, because optical continuum and emission lines originate in different regions.
Furthermore, this change does not appear to be connected with an increased luminosity in $\gamma$ rays, suggesting that the bluer continuum does not originate in the jet. 
Another hint of this is that these two points in Fig.~\ref{fig:corr_ew_cont} show a different behavior with respect to the others. 
It is therefore possible that such variation is due to a change of the physical properties in the accretion disk, and particularly in the accretion flow. 
The continuum luminosity indeed is often associated with the Eddington ratio \citep[e.g.][]{Wandel91}.
If we interpret the increased continuum luminosity visible in Fig.~\ref{fig:spind} as an increased Eddington ratio, this might in turn translate into a higher accretion rate onto the black hole. 
This possibility is consistent with the physical parameters derived for the flare. 
The strong magnetic field measured in the low state might hamper the accretion rate by moving plasma from the accretion disk to the jet base. 
During the outburst, when the energy is released in the form of a plasma condensation, the magnetic field suddenly decreases.
The accretion onto the black hole, therefore, is not suppressed anymore, and it might temporarily increase, until the magnetic field lines are fully restored. 
Such an efficient accretion could increase the disk temperature, leading then to the bluer continuum we observed. 
Unfortunately, our data can only provide a marginal evidence for this scenario. 
Optical spectra obtained during outburst events are indeed necessary to fully establish the behavior of the accretion disk.

\section{Summary}

In this paper we studied the multiwavelength behavior of a flat-spectrum radio quasar, 3C 345. 
In particular, we analyzed its evolution from the high state, in which its relativistic jet activity was very strong, to a very low state. 
The radio, optical, and $\gamma$-ray light curves allowed us to establish that at least three flaring episodes occurred between 2008 August and 2016 September, namely 2009A, 2009B, and 2015. 
In particular, our data allowed us to obtain a multiwavelength spectral energy distribution (SED) of the 2009B outburst, along with that of the low state of 2016 June. 
The SEDs indicated that the source became harder during the outburst, and that the magnetic field of the jet strongly decreased.
The analysis of several optical spectra observed in different epochs might indicate that the origin of the $\gamma$-ray emission is located close to the central engine and within the BLR. 
Moreover, the slope of the spectra allowed us to determine that the decrease in the magnetic field leads to an increase in the accretion, and to a bluer continuum. 

More multiwavelength observations of this and other sources, especially optical spectra, are needed to better constrain their behavior during flares, and to better understand how disk and jet interact with each other. 
In particular, the site from which the $\gamma$-ray flaring activity originates is still poorly understood, since the analysis of several events apparently suggests different scenarios. 
It has been suggested that IR photons originated in the dusty obscuring regions, surrounding the AGN on larger scales, make an important contribution to the production of $\gamma$ rays by means of IC processes. 
Such an interpretation, however, encounters more difficulties for this particular case. 
More data in the region between $\sim$100 keV and $\sim$100 MeV, which is still relatively unexplored and could allow us to understand whether breaks or more emission components are present in the IC hump, are crucial to address this issue. 
New facilities, such as the next generation telescopes e-ASTROGAM or AMEGO, could provide this kind of information, and improve our knowledge of the physical processes occurring in FSRQs.

\begin{acknowledgements}
We thank our referee Kirk Korista for comments which improved the quality of the paper. The authors are also grateful to Paola Marziani for providing helpful suggestions about the optical spectra analysis. We thank the internal referee of the \textit{Fermi LAT} Collaboration Stefano Ciprini and the \textit{Fermi LAT} coordinators Marcello Giroletti and Sara Cutini for a critical review of the paper. We thank the students of the 2013 NEON observing school Benjamin Hendricks, Jaan Laur, Maijana Smailagi\'c, and Sonia Tamburri, who performed the observations of 3C 345 in July 2013. N.H.L. acknowledges the support of the National Natural Science Foundation of China under grant 11703093. P.C. acknowledges funds from a 2017 "Research and Education" grant from Fondazione CRT". This work has been partially supported by PRIN INAF 2014 ``Jet and astro-particle physics of $\gamma$-ray blazars'' (P.I. F. Tavecchio). This work is based on observations made with the Copernico and Schmidt Telescopes of the INAF-Asiago Observatory, and the Galileo 1.22m telescope of the Asiago Astrophysical Observatory operated by the Department of Physics and Astronomy "G. Galilei" of the University of Padova. The Astronomical Observatory of the Autonomous Region of the Aosta Valley (OAVdA) is managed by the Fondazione Cl\'ement Fillietroz-ONLUS, which is supported by the Regional Government of the Aosta Valley, the Town Municipality of Nus and the "Unit\'e des Communes valdôtaines Mont-\'Emilius". We acknowledge the support of the staff of the Lijiang 2.4 m telescope and Kunming 1 m telescope. Funding for these telescopes has been provided by CAS and the People’s Government of Yunnan Province. The CSS survey is funded by the National Aeronautics and Space Administration under Grant No. NNG05GF22G issued through the Science Mission Directorate Near-Earth Objects Observations Program. The CRTS survey is supported by the U.S.~National Science Foundation under grants AST-0909182 and AST-1313422. Data from the Steward Observatory spectropolarimetric monitoring project were used. This program is supported by \textit{Fermi} Guest Investigator grants NNX08AW56G, NNX09AU10G, NNX12AO93G, and NNX15AU81G. The OVRO 40 M Telescope \textit{Fermi} Blazar Monitoring Program is supported by NASA under awards NNX08AW31G and NNX11A043G, and by the NSF under awards AST-0808050 and AST-1109911. This research has made use of the NASA/IPAC Extragalactic Database (NED) which is operated by the Jet Propulsion Laboratory, California Institute of Technology, under contract with the National Aeronautics and  Space Administration. Funding for the Sloan Digital Sky Survey has been provided by the Alfred P. Sloan Foundation, and the U.S. Department of Energy Office of Science. The SDSS web site is \texttt{http://www.sdss.org}. SDSS-III is managed by the Astrophysical Research Consortium for the Participating Institutions of the SDSS-III Collaboration including the University of Arizona, the Brazilian Participation Group, Brookhaven National Laboratory, Carnegie Mellon University, University of Florida, the French Participation Group, the German Participation Group, Harvard University, the Instituto de Astrofisica de Canarias, the Michigan State/Notre Dame/JINA Participation Group, Johns Hopkins University, Lawrence Berkeley National Laboratory, Max Planck Institute for Astrophysics, Max Planck Institute for Extraterrestrial Physics, New Mexico State University, University of Portsmouth, Princeton University, the Spanish Participation Group, University of Tokyo, University of Utah, Vanderbilt University, University of Virginia, University of Washington, and Yale University. This research has made use of the XRT Data Analysis Software (XRTDAS) developed under the responsibility of the ASI Science Data Center (ASDC), Italy. The \textit{Fermi} LAT Collaboration acknowledges generous ongoing support from a number of agencies and institutes that have supported both the  development and the operation of the LAT as well as scientific data analysis.These include the National Aeronautics and Space Administration and the Department of Energy in the United States, the Commissariat \`a l'Energie Atomique and the Centre National de la Recherche Scientifique / Institut National de Physique Nucl\'eaire et de Physique des Particules in France, the Agenzia Spaziale Italiana and the Istituto Nazionale di Fisica Nucleare in Italy, the Ministry
of Education, Culture, Sports, Science and Technology (MEXT), High Energy Accelerator Research Organization (KEK) and Japan Aerospace Exploration Agency (JAXA) in Japan, and the K.~A.~Wallenberg Foundation, the Swedish Research Council and the Swedish National Space Board in Sweden. Additional support for science analysis during the operations phase is gratefully acknowledged from the Istituto Nazionale di Astrofisica in Italy and the Centre National d'\'Etudes Spatiales in France. This work performed in part under DOE Contract DE-AC02-76SF00515.

\end{acknowledgements}

\bibliographystyle{aa}
\bibliography{/home/berton/Scrivania/Paper/biblio}

\clearpage
\begin{appendix}
\section{Online tables}
\begin{table}[h!]
\caption{X-ray parameters measured by Swift/XRT.}
\label{tab:X_flux}
\centering
\scalebox{0.9}{
\begin{tabular}{l c c c c}
\hline\hline
ObsID & Date & Flux & Photon Index & HR\\
\hline\hline
00036390002 & 2008-11-09 & 5.16$\pm$0.36 & 1.70$^{+0.20}_{-0.20}$ & -0.12$\pm$0.02 \\
00090085001 & 2009-05-12 & 8.11$\pm$0.48 & 1.64$^{+0.16}_{-0.16}$ & -0.54$\pm$0.06 \\
00090085002 & 2009-06-09 & 7.34$\pm$0.37 & 1.75$^{+0.15}_{-0.14}$ & -0.61$\pm$0.05 \\
00036390003 & 2009-07-26 & 6.55$\pm$0.26 & 1.70$^{+0.10}_{-0.10}$ & -0.58$\pm$0.04 \\
00036390004 & 2009-08-09 & 5.88$\pm$0.25 & 1.72$^{+0.10}_{-0.10}$ & -0.59$\pm$0.04 \\
00036390005 & 2009-08-23 & 5.80$\pm$0.29 & 1.68$^{+0.14}_{-0.13}$ & -0.56$\pm$0.05 \\
00036390006 & 2009-09-11 & 5.63$\pm$0.23 & 1.79$^{+0.10}_{-0.10}$ & -0.63$\pm$0.04 \\
00036390007 & 2009-09-20 & 6.15$\pm$0.35 & 1.59$^{+0.18}_{-0.17}$ & -0.51$\pm$0.05 \\
00036390008 & 2009-10-04 & 6.55$\pm$0.27 & 1.61$^{+0.11}_{-0.10}$ & -0.52$\pm$0.04 \\
00036390009 & 2009-10-06 & 6.43$\pm$0.51 & 1.51$^{+0.23}_{-0.22}$ & -0.46$\pm$0.05 \\
00036390010 & 2009-10-08 & 7.05$\pm$0.41 & 1.66$^{+0.17}_{-0.16}$ & -0.55$\pm$0.06 \\
00036390011 & 2009-10-09 & 7.01$\pm$0.42 & 1.52$^{+0.15}_{-0.15}$ & -0.46$\pm$0.05 \\
00036390012 & 2009-10-18 & 5.84$\pm$0.27 & 1.60$^{+0.12}_{-0.12}$ & -0.51$\pm$0.04 \\
00036390013 & 2009-11-01 & 5.15$\pm$0.23 & 1.69$^{+0.12}_{-0.11}$ & -0.57$\pm$0.04\\
00036390014 & 2009-11-15 & 5.44$\pm$0.24 & 1.80$^{+0.11}_{-0.11}$ & -0.63$\pm$0.04 \\
00036390015 & 2009-11-29 & 5.45$\pm$0.24 & 1.62$^{+0.13}_{-0.13}$ & -0.53$\pm$0.04 \\
00036390016 & 2010-03-06 & 4.67$\pm$0.60 & 1.85$^{+0.54}_{-0.50}$ & -0.66$\pm$0.13 \\
00036390017 & 2010-03-06 & 5.71$\pm$0.72 & 1.69$^{+0.30}_{-0.30}$ & -0.50$\pm$0.11 \\
00036390018 & 2010-03-07 & 4.49$\pm$0.41 & 1.94$^{+0.31}_{-0.29}$ & -0.70$\pm$0.09 \\
00036390019 & 2010-03-09 & 6.26$\pm$0.56 & 1.38$^{+0.38}_{-0.38}$ & -0.36$\pm$0.07 \\
00036390020 & 2010-04-09 & 5.16$\pm$0.35 & 1.73$^{+0.22}_{-0.21}$ & -0.60$\pm$0.06 \\
00036390021 & 2010-08-18 & 4.74$\pm$0.23 & 1.58$^{+0.14}_{-0.14}$ & -0.50$\pm$0.04 \\
00036390022 & 2010-11-01 & 4.94$\pm$0.32 & 1.58$^{+0.17}_{-0.17}$ & -0.50$\pm$0.06 \\
00091091001 & 2011-04-05 & 2.93$\pm$0.48 & 2.12$^{+0.72}_{-0.69}$ & -0.77$\pm$0.13 \\
00091091002 & 2011-04-05 & 4.56$\pm$0.37 & 1.55$^{+0.22}_{-0.21}$ & -0.48$\pm$0.07 \\
00036390023 & 2012-04-11 & 3.22$\pm$0.42 & 2.06$^{+0.50}_{-0.45}$ & -0.75$\pm$0.13 \\
00036390024 & 2013-05-26 & 3.51$\pm$0.57 & 1.35$^{+0.41}_{-0.41}$ & -0.41$\pm$0.14 \\
00091896001 & 2014-08-15 & 3.08$\pm$0.22 & 1.50$^{+0.19}_{-0.18}$ & -0.45$\pm$0.06 \\
00091896002 & 2014-09-15 & 2.24$\pm$0.17 & 1.76$^{+0.21}_{-0.21}$ & -0.61$\pm$0.07 \\
00091896003 & 2014-10-15 & 3.18$\pm$0.23 & 1.49$^{+0.21}_{-0.20}$ & -0.44$\pm$0.06 \\
00036390025 & 2016-04-25 & 3.73$\pm$0.34 & 1.42$^{+0.40}_{-0.40}$ & -0.39$\pm$0.09 \\
00036390026 & 2016-05-27 & 4.59$\pm$0.59 & 1.40$^{+0.84}_{-0.80}$ & -0.38$\pm$0.11 \\
00036390027 & 2016-06-23 & 3.00$\pm$0.27 & 1.31$^{+0.17}_{-0.17}$ & -0.30$\pm$0.07 \\
\hline\hline
\end{tabular}
}
\tablefoot{Columns: (1) Observation ID; (2) observation date; (3) flux 0.3-10 keV ($\times10^{-12}$ erg s$^{-1}$ cm$^{-2}$); (4) photon index with errors; (5) hardness ratio.}
\end{table}

\begin{table}[ht!]
\caption{Parameters of the optical spectra.}
\label{tab:spectral_parameters}
\centering
\scalebox{0.9}{
\begin{tabular}{l c c c c}
\hline\hline
Date & $\log \lambda$L$_{3000}$ & $\log$L$_{Mg \, II}$ & $\alpha_\lambda$ & EW (Mg II) \\
\hline\hline
2009-03-24 & 41.89$\pm$0.42 & 43.54$\pm$0.02 & -0.93$\pm$0.06 & 48.8$\pm$1.1 \\
2009-08-20 & 42.00$\pm$0.42 & 43.37$\pm$0.03 & -0.83$\pm$0.08 & 35.1$\pm$0.8 \\
2009-09-15 & 42.20$\pm$0.42 & 43.59$\pm$0.01 & -0.86$\pm$0.05 & 27.0$\pm$0.6 \\
2009-09-16 & 42.21$\pm$0.42 & 43.54$\pm$0.02 & -0.84$\pm$0.05 & 29.1$\pm$0.6 \\
2009-09-17 & 42.23$\pm$0.42 & 43.58$\pm$0.01 & -0.94$\pm$0.06 & 30.5$\pm$0.7 \\
2009-09-18 & 42.23$\pm$0.42 & 43.57$\pm$0.02 & -0.77$\pm$0.03 & 25.0$\pm$0.5 \\
2009-09-19 & 42.25$\pm$0.42 & 43.59$\pm$0.02 & -0.82$\pm$0.04 & 25.4$\pm$0.6 \\
2009-09-20 & 42.28$\pm$0.42 & 43.59$\pm$0.01 & -0.88$\pm$0.04 & 22.6$\pm$0.5 \\
2009-10-19 & 41.94$\pm$0.42 & 43.59$\pm$0.02 & -1.12$\pm$0.12 & 51.5$\pm$1.1 \\
2009-10-20 & 41.92$\pm$0.42 & 43.57$\pm$0.01 & -0.99$\pm$0.06 & 58.7$\pm$1.3 \\
2009-10-21 & 41.90$\pm$0.42 & 43.54$\pm$0.03 & -0.87$\pm$0.06 & 58.7$\pm$1.3 \\
2009-10-22 & 41.91$\pm$0.42 & 43.57$\pm$0.02 & -0.95$\pm$0.05 & 58.8$\pm$1.3 \\
2009-10-23 & 41.88$\pm$0.42 & 43.56$\pm$0.02 & -0.88$\pm$0.08 & 59.7$\pm$1.3 \\
2009-10-24 & 41.89$\pm$0.42 & 43.57$\pm$0.02 & -1.00$\pm$0.07 & 61.3$\pm$1.3 \\
2009-10-25 & 41.88$\pm$0.42 & 43.57$\pm$0.01 & -1.07$\pm$0.07 & 59.8$\pm$1.3 \\
2009-10-26 & 41.88$\pm$0.42 & 43.50$\pm$0.02 & -0.92$\pm$0.06 & 59.6$\pm$1.3 \\
2009-11-17 & 42.13$\pm$0.42 & 43.88$\pm$0.02 & -1.44$\pm$0.16 & 65.0$\pm$1.4 \\
2009-11-19 & 41.87$\pm$0.42 & 43.65$\pm$0.03 & -1.91$\pm$0.24 & 90.9$\pm$2.0 \\
2010-02-13 & 41.81$\pm$0.42 & 43.54$\pm$0.02 & -1.09$\pm$0.12 & 73.3$\pm$1.6 \\
2010-02-14 & 41.85$\pm$0.42 & 43.55$\pm$0.01 & -1.14$\pm$0.08 & 65.8$\pm$1.4 \\
2010-02-17 & 41.85$\pm$0.42 & 43.59$\pm$0.01 & -1.23$\pm$0.06 & 70.6$\pm$1.5 \\
2010-03-16 & 41.80$\pm$0.42 & 43.56$\pm$0.01 & -1.07$\pm$0.04 & 67.7$\pm$1.5 \\
2010-03-19 & 41.83$\pm$0.42 & 43.55$\pm$0.01 & -0.99$\pm$0.04 & 66.7$\pm$1.4 \\
2010-04-06 & 41.99$\pm$0.42 & 43.57$\pm$0.01 & -0.89$\pm$0.03 & 49.7$\pm$1.1 \\
2010-04-08 & 41.98$\pm$0.42 & 43.56$\pm$0.01 & -0.91$\pm$0.05 & 46.4$\pm$1.0 \\
2010-04-09 & 42.01$\pm$0.42 & 43.56$\pm$0.01 & -0.80$\pm$0.07 & 45.9$\pm$1.0 \\
2010-04-11 & 41.97$\pm$0.42 & 43.55$\pm$0.01 & -0.80$\pm$0.04 & 45.5$\pm$1.0 \\
2010-05-15 & 41.91$\pm$0.42 & 43.65$\pm$0.01 & -1.15$\pm$0.08 & 72.4$\pm$1.6 \\
2010-05-16 & 41.82$\pm$0.42 & 43.57$\pm$0.01 & -1.06$\pm$0.09 & 70.1$\pm$1.5 \\
2010-06-11 & 41.84$\pm$0.42 & 43.56$\pm$0.01 & -1.12$\pm$0.03 & 74.9$\pm$1.6 \\
2010-06-13 & 41.85$\pm$0.42 & 43.56$\pm$0.02 & -1.06$\pm$0.09 & 63.9$\pm$1.4 \\
2010-06-16 & 41.87$\pm$0.42 & 43.57$\pm$0.02 & -1.13$\pm$0.03 & 66.8$\pm$1.5 \\
2010-06-30 & 41.85$\pm$0.42 & 43.52$\pm$0.03 & -1.29$\pm$0.09 & 71.4$\pm$1.6 \\
2010-07-10 & 41.97$\pm$0.42 & 43.66$\pm$0.01 & -0.88$\pm$0.09 & 63.8$\pm$1.4 \\
2010-09-03 & 41.76$\pm$0.42 & 43.59$\pm$0.01 & -1.21$\pm$0.02 & 80.5$\pm$1.7 \\
2010-09-05 & 41.75$\pm$0.42 & 43.55$\pm$0.01 & -1.09$\pm$0.04 & 83.6$\pm$1.8 \\
2010-10-07 & 41.72$\pm$0.42 & 43.57$\pm$0.02 & -1.17$\pm$0.06 & 88.5$\pm$1.9 \\
2010-10-08 & 41.75$\pm$0.42 & 43.59$\pm$0.01 & -1.27$\pm$0.07 & 89.9$\pm$2.0 \\
2011-02-06 & 41.72$\pm$0.42 & 43.54$\pm$0.01 & -1.19$\pm$0.06 & 75.0$\pm$1.6 \\
2011-03-05 & 41.78$\pm$0.42 & 43.61$\pm$0.01 & -1.04$\pm$0.10 & 88.4$\pm$1.9 \\
2011-04-08 & 41.68$\pm$0.42 & 43.55$\pm$0.01 & -1.06$\pm$0.07 & 98.2$\pm$2.1 \\
2011-05-27 & 41.68$\pm$0.42 & 43.57$\pm$0.01 & -1.25$\pm$0.11 & 91.1$\pm$2.0 \\
2011-05-29 & 41.75$\pm$0.42 & 43.68$\pm$0.02 & -1.26$\pm$0.05 & 87.9$\pm$1.9 \\
2011-07-01 & 41.63$\pm$0.42 & 43.47$\pm$0.02 & -1.21$\pm$0.16 & 90.0$\pm$2.0 \\
2011-07-30 & 41.39$\pm$0.42 & 43.22$\pm$0.03 & -1.38$\pm$0.09 & 106.0$\pm$2.3 \\
2011-09-28 & 41.63$\pm$0.42 & 43.57$\pm$0.01 & -1.10$\pm$0.10 & 96.3$\pm$2.1 \\
2011-10-03 & 41.63$\pm$0.42 & 43.55$\pm$0.01 & -1.20$\pm$0.10 & 97.3$\pm$2.1 \\
2012-03-26 & 41.57$\pm$0.42 & 43.54$\pm$0.01 & -1.12$\pm$0.12 & 109.8$\pm$2.4 \\
2012-05-28 & 41.67$\pm$0.42 & 43.54$\pm$0.01 & -1.02$\pm$0.09 & 82.0$\pm$1.8 \\
2012-06-12 & 41.68$\pm$0.42 & 43.44$\pm$0.02 & -1.30$\pm$0.07 & 77.9$\pm$1.7 \\
2012-06-17 & 41.65$\pm$0.42 & 43.51$\pm$0.01 & -0.84$\pm$0.05 & 88.7$\pm$1.9 \\
2012-09-15 & 41.56$\pm$0.42 & 43.47$\pm$0.01 & -0.93$\pm$0.04 & 99.6$\pm$2.2 \\
2013-07-16 & 41.45$\pm$0.43 & 43.46$\pm$0.02 & -1.06$\pm$0.21 & 127.9$\pm$2.8 \\
2016-05-28 & 41.41$\pm$0.42 & 43.44$\pm$0.01 & -1.02$\pm$0.08 & 129.0$\pm$2.8 \\
\hline\hline
\end{tabular}
}
\tablefoot{Columns: (1) Observation date; (2) logarithm of the monochromatic continuum luminosity at 3000$\AA{}$ (erg s$^{-1}$); (3) logarithm of the total Mg II luminosity; (4) spectral index, defined as $F_\lambda \propto \lambda^{\alpha_\lambda}$; (5) equivalent width of Mg II.}
\end{table}

\begin{sidewaystable*}
\caption{Observed optical magnitudes from different instruments.}
\label{tab:mag_uv}
\centering
\begin{tabular}{l c c c c c c c c c c c c c c c}
\hline\hline
Date & Telescope & U & B & V & R & I & g & r & i & uvw1 & uvm2 & uvw2\\
\hline\hline
2008-11-09 & UVOT & 16.98$\pm$0.03 & {} & {} & {} & {} & {} & {} & {}  & 16.70$\pm$0.04 & {} & {} \\
2009-05-12 & UVOT & {} & {} & {} & {} & {} & {} & {} & {} & 16.35$\pm$0.03 & {} & {}\\
2009-06-09 & UVOT & {} & {} & {} & {} & {} & {} & {} & {} & 16.50$\pm$0.03 & {} & {}\\
2009-07-26 & UVOT & 16.84$\pm$0.04 & 17.49$\pm$0.04 & 17.28$\pm$0.06 & {} & {} & {} & {} & {}  & 16.47$\pm$0.04 & 16.42$\pm$0.04 & 16.50$\pm$0.03\\
2009-08-09 & UVOT & 16.73$\pm$0.04 & 17.43$\pm$0.04 & 17.11$\pm$0.06 & {} & {} & {} & {} & {}  & 16.36$\pm$0.03 & 16.33$\pm$0.04 & 16.40$\pm$0.03\\
2009-08-23 & UVOT & 16.67$\pm$0.05 & 17.38$\pm$0.05 & 17.00$\pm$0.06 & {} & {} & {} & {} & {}  & 16.39$\pm$0.04 & 16.43$\pm$0.04 & 16.48$\pm$0.03 \\
2009-09-11 & UVOT & 16.23$\pm$0.04 & 16.98$\pm$0.04 & 16.55$\pm$0.04 & {} & {} & {} & {} & {}  & 16.07$\pm$0.03 & 16.07$\pm$0.03 & 16.14$\pm$0.03\\
2009-10-04 & UVOT & 16.05$\pm$0.03 & 16.80$\pm$0.03 & 16.47$\pm$0.04 & {} & {} & {} & {} & {}  & 15.90$\pm$0.03 & 15.89$\pm$0.03 & 16.01$\pm$0.03\\
2009-10-06 & UVOT & 16.42$\pm$0.05 & 17.04$\pm$0.04 & 16.76$\pm$0.07 & {} & {} & {} & {} & {}  & 16.21$\pm$0.04 & 16.15$\pm$0.04 & 16.29$\pm$0.03\\
2009-10-08 & UVOT & 16.48$\pm$0.04 & 17.12$\pm$0.04  & 16.87$\pm$0.07 & {} & {} & {} & {} & {}  & 16.26$\pm$0.04 & 16.19$\pm$0.05 & 16.25$\pm$0.03\\
2009-10-09 & UVOT & 16.53$\pm$0.04 & 17.19$\pm$0.04 & 16.91$\pm$0.08 & {} & {} & {} & {} & {}  & 16.26$\pm$0.04 & 16.13$\pm$0.05 & 16.32$\pm$0.03\\
2009-10-18 & UVOT & 16.54$\pm$0.04 & 17.32$\pm$0.04 & 17.04$\pm$0.06 & {} & {} & {} & {} & {}  & 16.29$\pm$0.03 & 16.23$\pm$0.04 &16.31$\pm$0.03 \\ 
2009-11-01 & UVOT & 16.56$\pm$0.04 & 17.27$\pm$0.04 & 17.06$\pm$0.06 & {} & {} & {} & {} & {}  & 16.30$\pm$0.03 & 16.22$\pm$0.04 & 16.28$\pm$0.03\\
2009-11-15 & UVOT & 16.79$\pm$0.04 & 17.62$\pm$0.04 & 17.34$\pm$0.07 & {} & {} & {} & {} & {}  & 16.40$\pm$0.03 & 16.30$\pm$0.04 & 16.37$\pm$0.03\\
2009-11-29 & UVOT & 16.89$\pm$0.04 & 17.63$\pm$0.04 & 17.43$\pm$0.07 & {} & {} & {} & {} & {}  & 16.45$\pm$0.04 & 16.39$\pm$0.04 & 16.43$\pm$0.03\\
2010-03-06 & UVOT & 16.78$\pm$0.06 & 17.38$\pm$0.06 & {} & {} & {} & {} & {} & {} & 16.43$\pm$0.05 & {} & 16.44$\pm$0.08\\
2010-03-06 & UVOT & {} & {} & {} & {} & {} & {} & {} & {} & {} & 16.34$\pm$0.05 & 16.39$\pm$0.04\\
2010-03-07 & UVOT & {} & {} & {} & {} & {} & {} & {} & {} & 16.45$\pm$0.03 & {} & {}\\
2010-03-08 & UVOT & 16.93$\pm$0.03 & {} & {} & {} & {} & {} & {} & {} & {} & {} & {}\\
2010-04-09 & UVOT & 16.44$\pm$0.05 & 17.16$\pm$0.04 & 16.86$\pm$0.08 & {} & {} & {} & {} & {} & 16.20$\pm$0.04 & 16.17$\pm$0.04 & 16.24$\pm$0.03\\
2010-06-30 & NOT & {} & {} & {} & 17.05$\pm$0.22 & {} & {} & {} & {} & {} & {} & {} \\
2010-08-17 & UVOT & 16.98$\pm$0.04 & 17.69$\pm$0.04 & 17.57$\pm$0.07 & {} & {} & {} & {} & {} & 16.54$\pm$0.03 & 16.44$\pm$0.04 & 16.56$\pm$0.03\\
2010-11-01 & UVOT & 16.86$\pm$0.05 & 17.73$\pm$0.06 & 17.41$\pm$0.10 & {} & {} & {} & {} & {} & 16.48$\pm$0.04 & 16.32$\pm$0.05 & 16.41$\pm$0.03\\
2011-04-05 & UVOT & 17.13$\pm$0.03 & {} & {} & {} & {} & {} & {} & {} & {} & {} & 16.65$\pm$0.05\\
2011-04-05 & UVOT & 17.11$\pm$0.08 & 17.99$\pm$0.10 & 17.50$\pm$0.17 & {} & {} & {} & {} & {} & 16.69$\pm$0.06 & 16.81$\pm$0.09 & 16.72$\pm$0.05\\
2011-05-30 & NOT & {} & {} & {} & 17.49$\pm$0.40 & {} & {} & {} & {}  & {} & {} & {}\\
2012-04-11 & UVOT & 17.38$\pm$0.10 & 18.05$\pm$0.10 & 17.74$\pm$0.18 & {} & {} & {} & {} & {} & 16.82$\pm$0.07 & 16.89$\pm$0.08 & 16.96$\pm$0.06\\
2012-07-15 & NOT & {} & 18.03$\pm$0.05 & 17.86$\pm$0.05 & 17.54$\pm$0.06 & {} & {} & {} & {} & {} & {} & {} \\
2013-05-26 & UVOT & 17.47$\pm$0.10 & 18.30$\pm$0.12 & 18.07$\pm$0.20 & {} & {} & {} & {} & {} & 17.33$\pm$0.09 & 17.31$\pm$0.10 & 17.38$\pm$0.07\\
2013-07-18 & NOT & {} & 18.39$\pm$0.23 & {} & 18.11$\pm$0.23 & {} & {} & {} & {}  & {} & {} & {}\\
2014-08-15 & UVOT & 17.70$\pm$0.03 & 18.42$\pm$0.10 & 18.65$\pm$0.23 & {} & {} & {} & {} & {} & 17.30$\pm$0.08 & 17.33$\pm$0.08 & 17.29$\pm$0.05\\
2014-09-15 & UVOT & 17.62$\pm$0.10 & 18.33$\pm$0.11 & 18.39$\pm$0.20 & {} & {} & {} & {} & {} & 17.47$\pm$0.03 & 17.35$\pm$0.08 & 17.30$\pm$0.06\\
2014-10-15 & UVOT & 17.72$\pm$0.11 & 18.31$\pm$0.11 & 18.11$\pm$0.19 & {} & {} & {} & {} & {} & 17.41$\pm$0.09 & 17.30$\pm$0.09 & 17.27$\pm$0.03\\
2016-04-16 & T182 & {} & {} & {} & {} & {} & 18.33$\pm$0.13 & 18.00$\pm$0.48 & 18.25$\pm$0.19  & {} & {} & {}\\
2016-04-25 & UVOT &  17.77$\pm$0.08 & 18.29$\pm$0.07 & 17.97$\pm$0.13 & {} & {} & {} & {} & {}  & 17.29$\pm$0.06 & 17.19$\pm$0.06 & 17.26$\pm$0.05\\
2016-05-27 & UVOT & 17.79$\pm$0.11 & 18.35$\pm$0.11 & 18.71$\pm$0.30 & {} & {} & {} & {} & {}  & 17.53$\pm$0.10 & 17.30$\pm$0.09 & 17.49$\pm$0.07\\
2016-06-16 & Schmidt & {} & 19.55$\pm$0.23 & 18.97$\pm$0.17 & 18.99$\pm$0.25 & 17.70$\pm$0.12 & {} & {} & {}  & {} & {} & {}\\
2016-06-26 & UVOT & 17.95$\pm$0.09 & 18.46$\pm$0.08 & 18.28$\pm$0.15 & {} & {} & {} & {} & {} & 17.53$\pm$0.07 & 17.50$\pm$0.08 & 17.45$\pm$0.05\\
2016-07-25 & Schmidt & {} & 18.59$\pm$0.17 & 18.42$\pm$0.10 & 18.09$\pm$0.23 & 17.29$\pm$0.29 & {} & {} & {}  & {} & {} & {}\\
\hline
\end{tabular}
\tablefoot{Columns: (1) Observation date; (2) telescope; (3) U Vega magnitude; (4) B Vega magnitude; (5) V Vega magnitude; (6) R magnitude; (7) I magnitude; (8) g magnitude; (9) r magnitude; (10) i magnitude.}
\end{sidewaystable*}

\clearpage

\end{appendix}
\end{document}